\documentclass[twocolumn,floatfix,preprintnumbers,nofootinbib,superscriptaddress]{revtex4}

\usepackage{ulem}
\usepackage{bm}
\usepackage{times}
\usepackage{amssymb,amsbsy,amsmath,amsfonts}
\usepackage{graphicx}
\usepackage{float}
\usepackage{color}
\usepackage{morefloats}
\usepackage{rotating}
\usepackage{srcltx}
\usepackage{slashed}
\usepackage{subfigure}
\usepackage{multirow}
\usepackage{verbatim}
\usepackage{hyperref}
\usepackage{tabularx}
\usepackage{braket}
\usepackage{diagbox}
\usepackage{appendix}
\usepackage[section]{placeins}

\usepackage[outdir=./]{epstopdf}

\begin{document}
\title{Faddeev fixed-center approximation to the $\eta K^*\bar{K}^*$, $\pi K^*\bar{K}^*$ and $KK^*\bar{K}^*$ systems}

 \author{Qing-Hua Shen}~\email{shenqinghua@impcas.ac.cn}
 \affiliation{Institute of Modern Physics, Chinese Academy of Sciences,Lanzhou 730000,China}
 \affiliation{School of Nuclear Sciences and Technology, University of Chinese Academy of Sciences,Beijing 101408, China}
 \author{Ju-Jun Xie}~\email{xiejujun@impcas.ac.cn}
 \affiliation{Institute of Modern Physics, Chinese Academy of Sciences,Lanzhou 730000,China}
 \affiliation{School of Nuclear Sciences and Technology, University of Chinese Academy of Sciences,Beijing 101408, China}
\affiliation{Lanzhou Center for Theoretical Physics, Key Laboratory of Theoretical Physics of Gansu Province, Lanzhou University, Lanzhou, Gansu 730000, China}

 \begin{abstract}

The three-body $\eta K^*\bar{K}^*$, $\pi K^*\bar{K}^*$ and $KK^*\bar{K}^*$ systems are investigated within the framework of fixed-center approximation to the Faddeev equations, where $K^*\bar{K}^*$ is treated as the scalar meson $f_0(1710)$. The interactions between $\pi$, $\eta$, $K$ and $K^*$ are taking from the chiral unitary approach. By scattering the $\eta$ meson on the clusterized $(K^*\bar{K}^*)_{f_0(1710)}$ system, we find a peak in the modulus squared of the three-body scattering amplitude and it can be associated as a bound state with quantum numbers $I^G(J^{PC})=0^+(0^{-+})$. Its mass and width are around 2054 MeV and 60 MeV, respectively. This state could be associated to the $\eta(2100)$ meson. For the $\pi (K^*\bar{K}^*)_{f_0(1710)}$ scattering, we find a bump structure around 1900-2000 MeV with quantum numbers $1^-({0^{-+}})$. While for the $K (K^*\bar{K}^*)_{f_0(1710)}$ system, there are three structures. One of them is much stable and its mass is about 2130 MeV. It is expected that these theoretical predictions here could be tested by future experimental measurements, such as by the BESIII, BelleII and LHCb collaborations.

 \end{abstract}
 
\date{\today}
 
 \maketitle

\section{Introduction}

It is known that quark models have achieved great success in studying the properties of hadrons, especially for these ground states. Within the quark models, it is commonly accepted that mesons are composed of quark and antiquark ($q\bar{q}$) and baryons are composed of three quarks $(qqq)$. Recently, the topic of meson-meson and meson-baryon states, with hadrons and hadrons governed by strong interactions, has been well developed by the combination of the chiral effective Lagrangians with nonperturbative unitary techniques in coupled channels, which has been a very fruitful scheme to study the nature of many hadronic states, both on light and heavy sectors~\cite{Oset:2016lyh,Guo:2017jvc,Dong:2021juy,Dong:2021bvy}. Some of them are not easily explained by the classical quark models.

Those hadronic states with exotic quantum numbers cannot be descried by quark models and they have more complex structures. For example, a state with exotic quantum numbers $J^{PC} = 1^{-+}$, denoted as $\eta_1(1855)$, is observed by the BESIII collaboration~\cite{BESIII:2022iwi,BESIII:2022riz}, which cannot be explained by traditional $q\bar{q}$ picture. However, it can be easily obtained in the molecular picture. In Refs.~\cite{Yang:2022lwq,Dong:2022cuw}, the $\eta_1(1855)$ is interpreted as a $\bar{K}K(1400)$ molecular state. In Ref.~\cite{Chen:2022qpd}, the $\eta_1(1855)$ state was investigated by the method of QCD sum rules and the results support
the interpretation of the $\eta_1(1855)$ as a $\bar{s}sg$ hybrid meson. On the other hand, those states with exotic quantum numbers can be dynamically generated from the three-body interactions. In Ref.~\cite{Zhang:2016bmy}, the meson $\pi_1(1600)$ with exotic quantum numbers $I^G(J^{PC}) = 1^-(1^{-+})$, is interpreted as a dynamically generated state of the $\pi K^*\bar{K}$ system by using the fixed-center approximation (FCA), where the two-body $\bar{K}K^*$ is fixed as $f_1(1285)$ state~\cite{Lutz:2003fm,Roca:2005nm,Zhou:2014ila,Geng:2015yta,Xie:2019iwz}. Similarly, in Ref.~\cite{Zhang:2019ykd}, possible exotic states with mass around $1700$ MeV and quantum numbers $I^G(J^{PC}) = 0^+(1^{-+})$ in the $\eta K^*\bar{K}$ system were predicted. 

Indeed, there is growing evidence that some existing and new observed hadronic states could be interpreted in terms of resonances or bound states of three hadrons~\cite{MartinezTorres:2020hus,Wu:2022ftm,Malabarba:2021taj}. Furthermore, some new states are also predicted in three-body systems~\cite{Luo:2021ggs,Luo:2022cun,Ikeno:2022jbb}. For example, $DKK$ and $DK\bar{K}$ systems were studied in Ref.~\cite{Debastiani:2017vhv} within FCA, where the evidence of a state with mass of about $2833$-$2858$ MeV was found in the $DK\bar{K}$ system. In Ref.~\cite{Ikeno:2022jbb}, $D^*D^*\bar{K}^*$ is studied by FCA, and they obtain bound states with isospin $I =1/2$ in total spin $J=0$, $1$, and $2$. These states are very narrow. In Ref.~\cite{Wu:2020job}, the $KD\bar{D}$ system is studied using the Gaussian expansion method, minimizing the energy of the system. A new hidden charm $K_c(4180)$ excited state was predicted. The same system is also studied in Ref.~\cite{Wei:2022jgc} within the framework of FCA and the results obtained are very similar. Besides, the FCA method have many achievements on the study of three-body systems~\cite{MartinezTorres:2008gy,Ren_2018,Dias:2017miz,Bayar:2011qj,Xiao:2011rc,Bayar:2013bta,Liang:2013yta,Durkaya:2015wra,Yamagata-Sekihara:2010muv,MartinezTorres:2018zbl,SanchezSanchez:2017xtl,Xie:2010ig,Xie:2011uw}. Those successes give us confidence in the FCA method that we use in the present work to study the $\eta K^*\bar{K}^*$, $\pi K^*\bar{K}^*$, and $K K^*\bar{K}^*$ systems. In the frame work of FCA, there is a cluster of two bound particles ($K^*\bar{K}^*$ in this work) and the third one ($\eta$, $\pi$, or $K$ in this work) collides with the components of this cluster without modifying its wave function. Certainly, if the third particle is lighter than the constituents of the cluster, the approximation is better. Note that the $\eta$, $\pi$, and $K$ mesons are much lighter than the $K^*/\bar{K}^*$ meson.

For the $K^*\bar{K}^*$ subsystem, it has strong attraction between $K^*$ and $\bar{K}^*$. In fact, the well established scalar meson $f_0(1710)$ was proposed to be a state dynamically generated from the vector meson-vector meson interactions in coupled channels~\cite{Geng:2008gx,Geng:2009gb,Du:2018gyn,Garcia-Recio:2013uva,Wang:2021jub}. Within this picture, the $f_0(1710)$ resonance can be interpreted as a $K^*\bar{K^*}$ molecular state, and the properties of $f_0(1710)$ resonance and its production have been investigated in Refs.~\cite{Nagahiro:2008bn,Branz:2009cv,Geng:2010kma,MartinezTorres:2012du,Xie:2014gla,Dai:2015cwa,Molina:2019wjj}.  

In this work, we investigate the three-body $\eta K^*\bar{K}^*$, $\pi K^*\bar{K}^*$, and $K K^*\bar{K}^*$ systems by considering the interactions of the three components among themselves, keeping in mind the expected strong correlations of the $K^*\bar{K}^*$ subsystem to make the $f_0(1710)$ state. Then in terms of two-body pseudoscalar meson-vector meson scattering amplitudes, which were obtained from the chiral unitary approach~\cite{Lutz:2003fm,Roca:2005nm}, we solve the Faddeev equations for the $\eta$, $\pi$, and $K$ scattering on the $K^*\bar{K}^*$ cluster in $S$-wave by using the fixed center approximation. In this way, it is quite probable to be able to generate pseudoscalar mesons with three-body nature. The above three-body systems have quantum numbers $J^{PC} = 0^{-+}$ and isospin $I=0$, $1$, and $1/2$, respectively.

The paper is organized as follows. In next section, we present the FCA method to the three-body $\eta K^*\bar{K}^*$, $\pi K^*\bar{K}^*$, and $K K^*\bar{K}^*$ systems. In Sec.~\ref{numericalresults}, our theoretical results and discussions are presented. Finally, a short summary is followed.

 \section{Formalism and ingredients} \label{formalism}
 
 We are interested in three different systems constituted by three mesons: $\eta K^*\bar{K}^*$, $\pi K^*\bar{K}^*$, and $KK^*\bar{K}^*$. In order to study the dynamics of these systems we will solve the Faddeev equations using the fixed-center approximation to obtain the total scattering amplitudes. In this section we will summarize the derivation of the three-body scattering amplitudes within the framework of the FCA.
 
 \subsection{Fixed-center approximation}
 
\begin{figure*}[htbp]
    \centering
\includegraphics[scale=0.6]{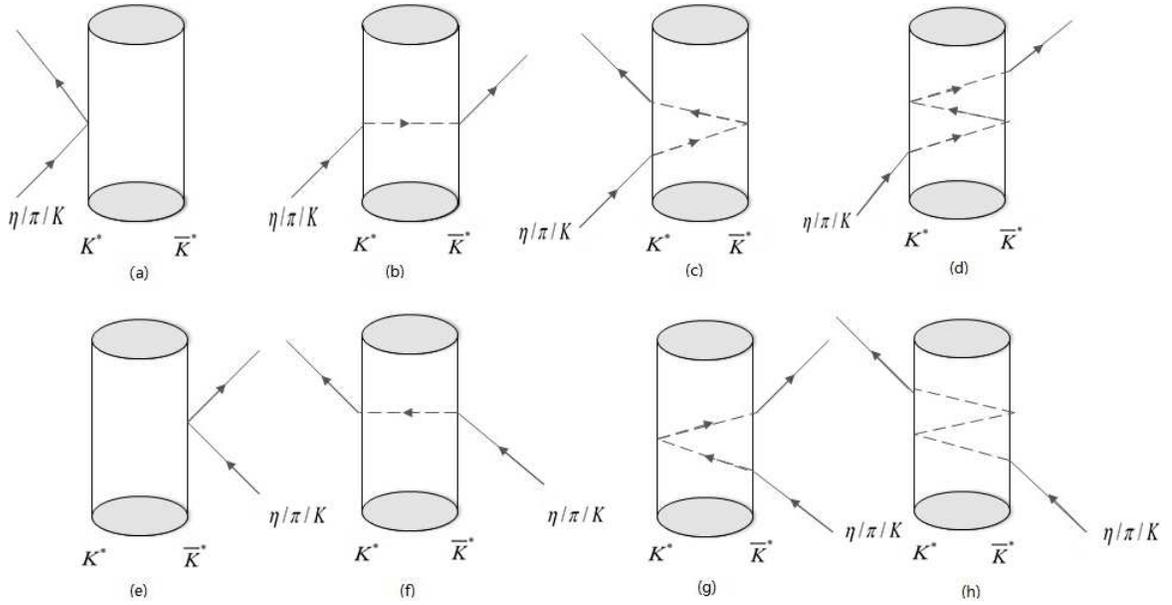}
\caption{Diagrammatic representation of FCA for the Faddeev equations for the three-body $\eta K^*\bar{K}^*$, $\pi K^*\bar{K}^*$, and $KK^*\bar{K}^*$ systems.}   \label{fig:FCA}
\end{figure*}

We are going to use the fixed-center approximation formalism to study the three-body $\eta K^*\bar{K}^*$, $\pi K^*\bar{K}^*$, and $KK^*\bar{K}^*$ systems. In this frame work, we consider the $K^*\bar{K}^*$ as a cluster, which we take it as the $f_0(1710)$ state, and $\eta$, $\pi$ or $K$ interacts with it. The corresponding diagrams are shown in Fig.~\ref{fig:FCA}. In the following we will label $K^*$, $\bar{K}^*$, and $\eta$ ($\pi$ or $K$) as particle 1, 2, and 3, respectively.

Following the formalism of Ref.~\cite{Roca:2010tf}, the three-body scattering amplitude $T$ for $\eta$ ($\pi$ or $K$) collisions with the $K^*\bar{K}^*$ cluster can be obtained by the sum of the partition functions $T_1$, and $T_2$:
\begin{eqnarray}
    T_1 &=& t_1+t_1G_0T_2,\\
    T_2 &=& t_2+t_2G_0T_1, \\
    T &=& T_1+T_2 = \frac{t_1 + t_2 +2t_1t_2G_0}{1-t_1t_2G^2_0},     \label{eq:T}
\end{eqnarray}
where $T_1$ stands for the sum of all the diagrams in Fig.~\ref{fig:FCA} where the third particle collides firstly with the particle $K^*$ in the cluster, while $T_2$ is the sum of all the diagrams where the third particle collides firstly with particle $\bar{K}^*$ of the cluster. The $t_1$ and $t_2$ represent the unitary two-body scattering amplitudes in coupled channels for the interactions of particle 3 with particle 1 and 2, respectively. These two-body scattering amplitudes will be discussed later on. 

Besides, in the above equations, $G_0$ is the loop function for the particle 3 propagating between the $K^*$ and $\bar{K}^*$ of the cluster, which can be written as,
\begin{align}
    G_0(s)=\frac{1}{2M_{f_0(1710)}}\int\frac{d^3\vec{q}}{(2\pi)^3}\frac{F_{f_0(1710)}(q)}{q^{0^2}-\left\lvert \vec{q}\!~ \right\rvert^2-m_3^2+i\epsilon },
\end{align}
where $s$ is the invariant mass square of the whole three-body system, and $M_{f_0(1710)}$ is the mass of the bound state $f_0(1710)$. Besides, $F_{f_0(1710)}(q)$ is the form factor of the $K^*\bar{K}^*$ subsystem, which is treated as the $f_0(1710)$ state. The $q^0$ is the energy of particle 3 with mass $m_3$ in the center-of-mass frame of particle 3 and the $(K^*\bar{K}^*)_{f_0(1710)}$ cluster, which is given by
\begin{align}
    \label{eq:q0}
    q^{0}(s)=\frac{s+m_3^2-M_{f_0(1710)}^2}{2\sqrt{s}}.
\end{align}

Following the approach of Refs.~\cite{Gamermann:2009uq,Yamagata-Sekihara:2010kpd,Roca:2010tf}, one can easily get the expression of the form factor $F_{f_0(1710)}$ for the $S$-wave $K^*\bar{K}^*$ bound state $f_0(1710)$ as
\begin{eqnarray}
F_{f_1(1710)}(q) &=& \frac{1}{N}\int_{\left\lvert \vec{p}\!~\right\rvert \leq \Lambda,\left\lvert \vec{p}-\vec{q}\!~\right\rvert\leq \Lambda  }d^3\vec{q}\frac{1}{2\omega_1(\vec{p}\!~)}\frac{1}{2\omega_2(\vec{p}\!~)} \times \nonumber \\ && \frac{1}{M_{f_0(1710)}-\omega_1(\vec{p}\!~)-\omega_2(\vec{p}\!~)} \times \nonumber \\ && \frac{1}{2\omega_1(\vec{p}-\vec{q}\!~)} \frac{1}{2\omega_2(\vec{p}-\vec{q}\!~)} \times \nonumber \\
&& \frac{1}{M_{f_1(1710)}-\omega_1(\vec{p}-\vec{q}\!~)-\omega_2(\vec{p}-\vec{q}\!~)}, \label{eqs:formfactor}
\end{eqnarray}
where $\omega_1(\vec{p}\!~) = \omega_2(\vec{p}\!~) = \sqrt{|\vec{p}\!~|^2+m^2_{K^*}}$, and the normalization factor $N$ is given by
\begin{equation}
N = \int_{\left\lvert \vec{p}\!~\right\rvert\leq \Lambda }d^3\vec{p}\left(\frac{1}{4\omega^2_1(\vec{p}\!~)}\frac{1}{M_{f_0(1710)}-2\omega_1(\vec{p}\!~)} \right)^2.
    \label{eqs:N}
\end{equation}
The cutoff parameter $\Lambda$ is needed to regularize the vector meson-vector meson loop functions in the chiral unitary approach~\cite{Geng:2008gx,Geng:2009gb}. In the present work, the upper integration limit of $\Lambda$ has the same value of the cutoff used in Refs.~\cite{Geng:2008gx,Geng:2009gb}. With the values of $\Lambda$, one can get the $f_0(1710)$ state in the vector meson-vector meson interactions in coupled channels.

\begin{figure}[htbp]
    \centering
    \includegraphics[scale=0.35]{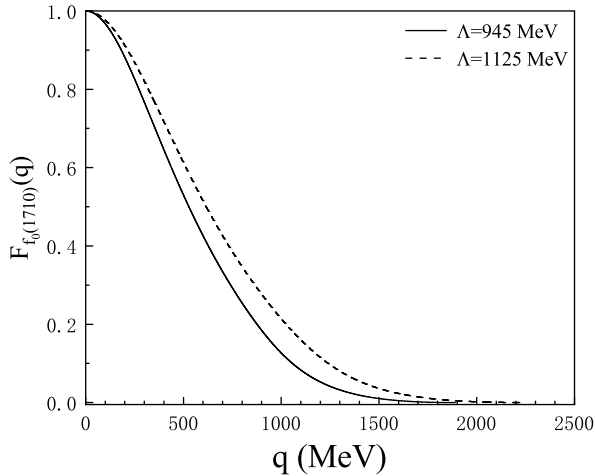}
    \caption{Form factor of $(K^*\bar{K}^*)_{f_0(1710)}$ cluster with $\Lambda=945$ MeV (solid line) and $\Lambda=1125$ MeV (dash line).}
    \label{fig:F}
\end{figure}

In Fig.~\ref{fig:F}, we show numerical results of the form factor $F_{f_0(1710)}$ as a function of $q = |\vec{q}\!~|$, where the solid and dash curves are obtained with $\Lambda = 945$ MeV and $1125$ MeV, respectively. Note that with these values for cutoff parameter $\Lambda$, one can get the $f_0(1710)$ state in the vector-vector interactions in coupled channels. Besides, the integration condition $|\vec{p} - \vec{q}\!~| < \Lambda$ implies that the form factor of $F_{f_0(1710)}$ is exactly zero when $q > 2\Lambda$.

\subsection{Unitarized $K^*\bar{K}^*$ interaction for the $f_0(1710)$ state and the loop function $G_0$}

\begin{figure}[htbp]
    \centering
    \includegraphics[scale=0.35]{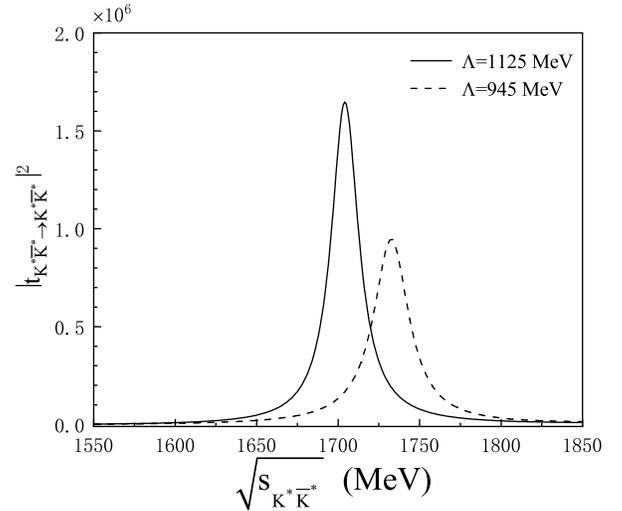}
    \caption{Modulus squared of $t_{K^* \bar{K}^* \to K^* \bar{K}^*}$ as a function of the invariant mass of $K^* \bar{K}^*$ system with $\Lambda=945$ MeV (dash line) and $\Lambda=1125$ MeV (solid line).}
    \label{fig:1710}
\end{figure}

The $K^*\bar{K}^*$ interaction in $S$-wave has been analyzed within the formalism developed in Ref.~\cite{Geng:2008gx} for studying the interaction of the nonet of vector mesons among themselves. In Fig.~\ref{fig:1710}, the module square of $t_{K^*\bar{K}^* \to K^*\bar{K}^*}$ obtained from the chiral unitary approach in $\rho \rho$, $\omega\omega$, $\omega \phi$, $\phi \phi$, and $K^*\bar{K}^*$ coupled channels for isospin $I=0$ sector, are shown, where one can see that there is a clear peak for $f_0(1710)$ state around $1704$ MeV with $\Lambda = 1125$ MeV. While the peak moves to $1733$ MeV if one take $\Lambda = 945$ MeV. Note that in the calculations of these two-body loop functions of $G_{\rho\rho}$ and $G_{K^*\bar{K}^*}$, the widths of vector mesons $\rho$ and $K^*$ are taken into account, since they have large total decay widths. On the other hand, in the calculations for the form factor of $F_{f_0(1710)}(q)$, the mass of cluster $f_0(1710)$ is taken as $1704$ MeV when $\Lambda = 1125$ MeV, while $M_{f_0(1710)}= 1733$ if one takes $\Lambda = 945$ MeV.

It is worthy to mention that the ``our evaluation” value in the 2020 version and 2021 updated of the Review of Particle Physics (RPP)~\cite{ParticleDataGroup:2020ssz} for the
mass of $f_0(1710)$ resonance is $1704 \pm 12$ MeV, while the ``our average" value for it is $1733^{+8}_{-7}$ MeV. In fact, even the $f_0(1710)$ resonance is well established in the RPP, there are still many doubts about its nature~\cite{Zhu:2022wzk}. We refer the Review of Particle Physics~\cite{ParticleDataGroup:2020ssz} for more details.

\begin{figure}[htbp]
    \centering
    \includegraphics[scale=0.35]{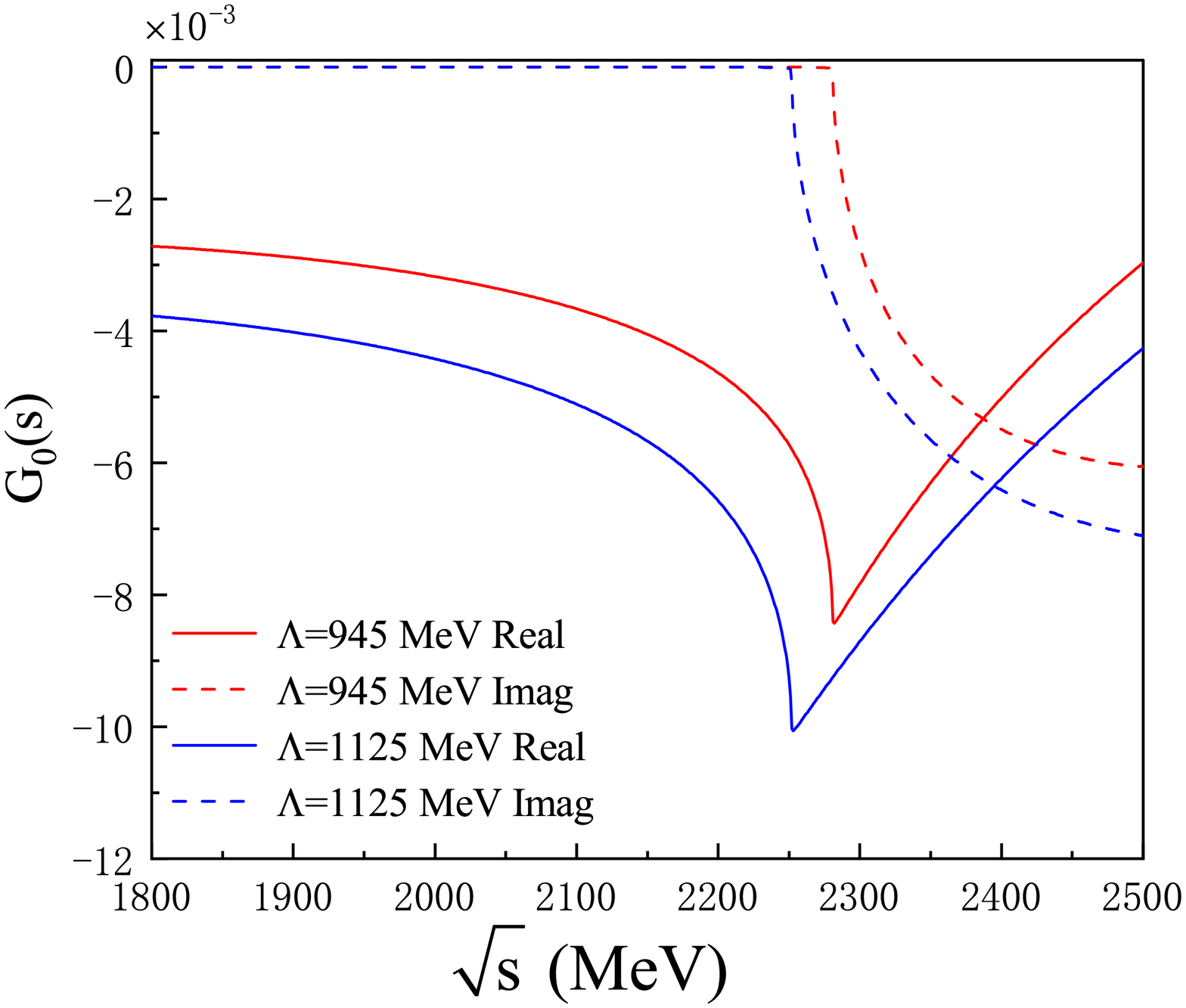}
    \caption{The real (solid line) and imaginary (dash line) parts of the loop function $G_0$ for $\eta K^*\bar{K}^*$ system with the cutoff parameter $\Lambda=945$ MeV (red line) and $\Lambda=1125$ MeV (blue line).}
    \label{fig:G0eta}
\end{figure}

\begin{figure}[htbp]
    \centering
    \includegraphics[scale=0.35]{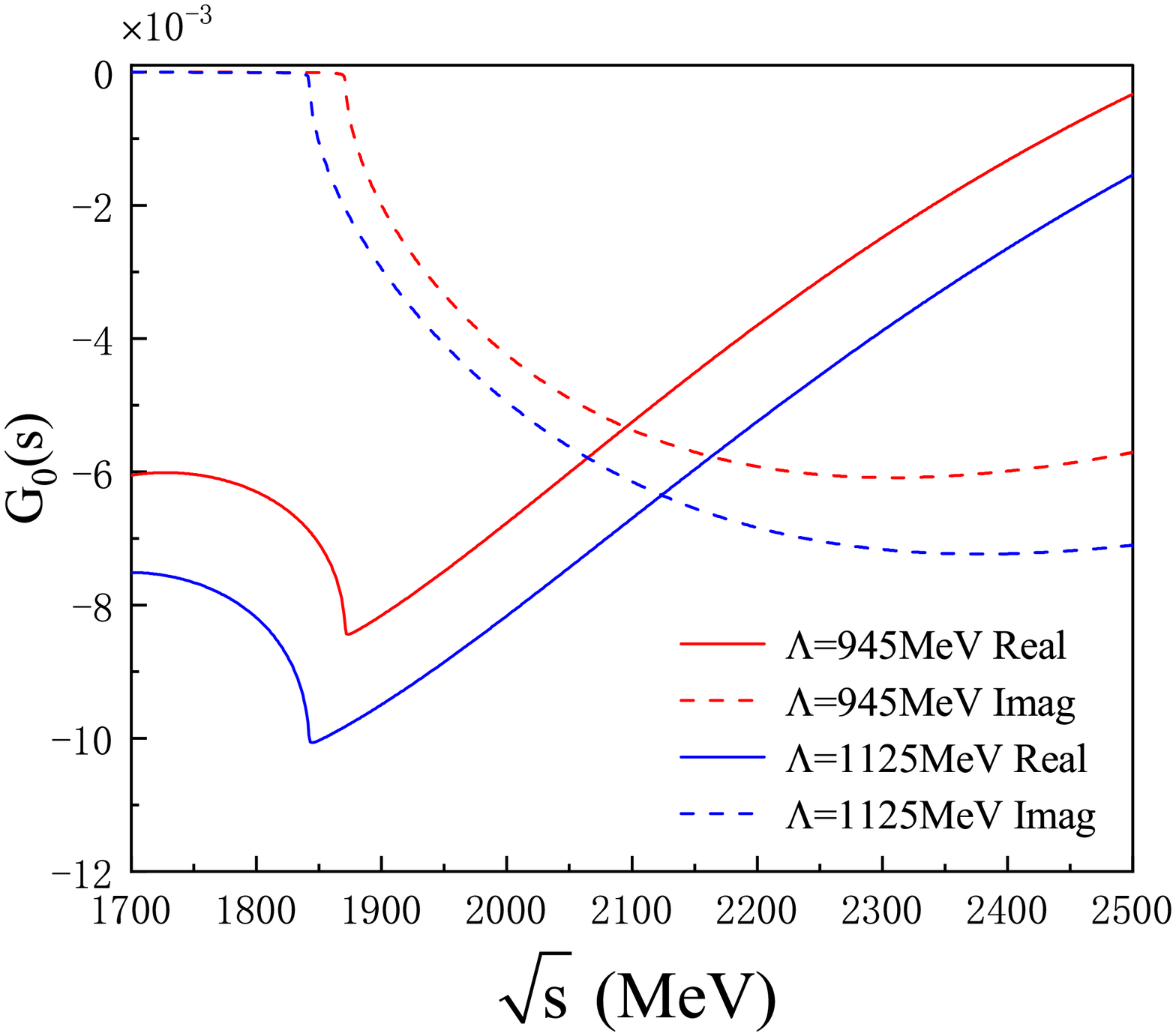}
    \caption{As in Fig.~\ref{fig:G0eta} but for the $\pi K^*\bar{K}^*$ system.}
    \label{fig:G0pi}
\end{figure}

\begin{figure}[htbp]
    \centering
    \includegraphics[scale=0.35]{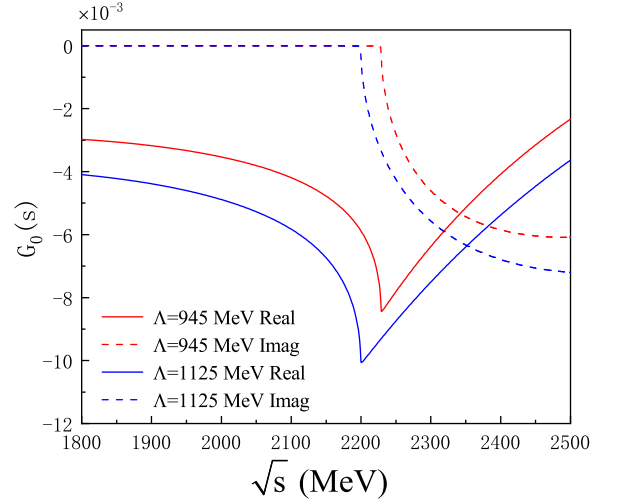}
    \caption{As in Fig.~\ref{fig:G0eta} but for the $K K^*\bar{K}^*$ system.}
    \label{fig:G0ka}
\end{figure}

Next we present the numerical results for the loop function $G_0$, which is the $\eta$ ($\pi$ and $K$) propagator between the $K^*$ and $\bar{K}^*$ of the cluster $f_0(1710)$. Note that $G_0$ is dependent on the invariant mass $\sqrt{s}$ of the three-body system. In this work, there are three $G_0$ functions for $\eta$-$(K^*\bar{K}^*)_{f_0(1710)}$, $\pi$-$(K^*\bar{K}^*)_{f_0(1710)}$, and $K$-$(K^*\bar{K}^*)_{f_0(1710)}$ systems. In Fig.~\ref{fig:G0eta}, we show the real and imaginary parts of the $G_0$ function for $I_{K^*\bar{K}^*} = 0$ as a function of the total three body $\eta$-$K^*$-$\bar{K}^*$ system invariant mass. The other two cases can be easily obtained by the replacements of $\eta \to \pi$ and $\eta \to K$, respectively. They are shown in Figs.~\ref{fig:G0pi} and \ref{fig:G0ka}.

\subsection{Contributions for the $\eta$, $\pi$, and $K$ interaction with the $K^*\bar{K}^*$ system}

According to Fig.~\ref{fig:FCA} (a) and (e), the single-scattering contribution from $t_1$ and $t_2$ are appropriate the combination of the two-body unitarized scattering amplitudes. For example, let us firstly consider the $\eta K^*\bar{K}^*$ system with the $K^*\bar{K}^*$ cluster in $I=0$ for the $f_0(1710)$ state. The $K^*\bar{K}^*$ isospin state is written as
\begin{align}
    \ket{K^*\bar{K}^*}_{I=0} = \frac{1}{\sqrt{2}}(\ket{\frac{1}{2},-\frac{1}{2}}-\ket{-\frac{1}{2},\frac{1}{2}}),
\end{align}
where the kets on the right side of the above equation represent $\ket{I_z^{K^*},I_z^{\bar{K}^*}}$ for $K^*\bar{K}^*$.

Then the single scattering contributions to the total amplitude of $\bra{\eta K^*\bar{K}^*}\hat{t}\ket{\eta K^*\bar{K}^*}$ can be easily obtained in terms of unitary two-body transition amplitudes $t_{\eta K^* \to \eta K^*}$ and $t_{\eta \bar{K}^* \to \eta \bar{K}^*}$ derived in Ref.~\cite{Roca:2005nm}. Here we write explicitly the case of $\eta (K^* \bar{K}^*)_{f_0(1710)} \to \eta (K^* \bar{K}^*)_{f_0(1710)}$,
\begin{equation}
    \begin{aligned}
        &\bra{\eta(K^*\bar{K}^*)}\hat{t}\ket{\eta(K^*\bar{K}^*)}\\
        &=(\bra{A_1}+\bra{A_2})(\hat{t}_{31}+\hat{t}_{32})(\ket{A_1}+\ket{A_2})\\
        &=\bra{A_1}\hat{t}_{31}\ket{A_1}+\bra{A_2}\hat{t}_{32}\ket{A_2},
    \end{aligned}
\end{equation}
where $\ket{A_1}$ stands for the state combined with $\eta$ and $K^*$, while $\ket{A_2}$ is the state of $\eta$ and $\bar{K}^*$. They are given by,
\begin{equation}
    \begin{aligned}
    \ket{A_1}  &=\frac{1}{\sqrt{2}}\ket{\frac{1}{2}\frac{1}{2},-\frac{1}{2}}-\frac{1}{\sqrt{2}}\ket{\frac{1}{2}-\frac{1}{2},\frac{1}{2}}, \\
     \ket{A_2} &=\frac{1}{\sqrt{2}}\ket{\frac{1}{2}-\frac{1}{2},\frac{1}{2}}-\frac{1}{\sqrt{2}}\ket{\frac{1}{2}\frac{1}{2},-\frac{1}{2}}.
    \end{aligned}
\end{equation}

Thus, we have
\begin{equation}
    \begin{aligned}
        t_1&=\bra{A_1}\hat{t}_{31}\ket{A_1}\\
       &=\frac{1}{2}t^{I=\frac{1}{2}}_{\eta K^*\rightarrow \eta K^* }+\frac{1}{2}t^{I=\frac{1}{2}}_{\eta K^*\rightarrow \eta K^* }, \\  
       &=t^{I=\frac{1}{2}}_{\eta K^*\rightarrow \eta K^* } \\
        t_2&=\bra{A_2}\hat{t}_{32}\ket{A_2}\\
        &=\frac{1}{2}t^{I=\frac{1}{2}}_{\eta \bar{K}^*\rightarrow \eta \bar{K}^*} + \frac{1}{2}t^{I=\frac{1}{2}}_{\eta \bar{K}^*\rightarrow \eta \bar{K}^* } \\ &=t^{I=\frac{1}{2}}_{\eta \bar{K}^*\rightarrow \eta \bar{K}^* }.
    \end{aligned}
\end{equation}.

Similarly, for the case of $\pi K^* \bar{K}^*$, we have
\begin{equation}
    \begin{aligned}
        t_1&=\bra{A_1}\hat{t}_{31}\ket{A_1}\\
         &=\frac{2}{3}t_{\pi K^*\rightarrow \pi K^*}^{I=\frac{3}{2}}+\frac{1}{3}t_{\pi K^*\rightarrow \pi K^*}^{I=\frac{1}{2}}, \\
        t_2 &=\bra{A_2}\hat{t}_{32}\ket{A_2}\\
        &=\frac{2}{3}t_{\pi \bar{K}^*\rightarrow \pi \bar{K}^*}^{I=\frac{3}{2}}+\frac{1}{3}t_{\pi \bar{K}^*\rightarrow \pi \bar{K}^*}^{I=\frac{1}{2}} .
    \end{aligned}
\end{equation}

Next, for the case of $K K^*\bar{K}^*$, we can get following results,
\begin{eqnarray}
    t_{1} &=&\bra{A_{1}}\hat{t_{31}}\ket{A_{1}} \nonumber \\
    &=&\frac{3}{4}t_{KK^* \to K K^*}^{I=1} + \frac{1}{4}t_{KK^* \to K K^*}^{I=0}, \\
    t_{2} &=& \bra{A_{2}}\hat{t_{32}}\ket{A_{32}} \nonumber \\
    &=&\frac{3}{4}t_{K\bar{K}^* \to K\bar{K}^*}^{I=1} + \frac{1}{4}t_{K\bar{K}^* \to K\bar{K}^*}^{I=0}.
\end{eqnarray}

On the other hand, following the approach developed in Refs.~\cite{Roca:2010tf,Yamagata-Sekihara:2010muv}, we need to give a weight to $t_1$ and $t_2$ such that we have the right normalization for the fields of mesons. This is achieved by replacing
\begin{equation}
    \begin{aligned}
        t_1 \to \tilde{t_1} &= \frac{M_{f_0(1710)}}{m_{K^*}}t_1, \\
       t_2 \to \tilde{t_2} &= \frac{M_{f_0(1710)}}{m_{\bar{K}^*}}t_2.
    \end{aligned}
\end{equation}

Finally, we have the three-body scattering amplitude:
\begin{align}
    T = \frac{\tilde{t_1}+\tilde{t_2}+2\tilde{t_1}\tilde{t_2}G_0}{1-\tilde{t_1}\tilde{t_2}G_0^2}.
\end{align}

It is worth to mention that the three-body total scattering amplitude $T$ is a function of the total invariant mass $\sqrt{s}$ of the three-body system. While the two-body scattering amplitudes $t_1$ and $t_2$ depend on the invariant masses $\sqrt{s_1}$ and $\sqrt{s_2}$, which are the invariant masses of $\eta$ ($\pi$ or $K$) and particle $K^*$ ($\bar{K}^*$) inside the cluster of $f_0(1710)$. The arguments $s_1$ and $s_2$ are obtained as
\begin{equation}
    \begin{aligned}
        s_1 &=m_{\eta/\pi/K}^2+m_{K^*}^2+\frac{s-m_{\eta/\pi /K}^2-M_{f_0(1710)}^2}{2}\\
        s_2 &= m_{\eta/\pi /K}^2+m_{\bar{K}^*}^2+\frac{s-m_{\eta/\pi/K}^2-M_{f_0(1710)}^2}{2}\\   
    \end{aligned}
\end{equation}
where we have used $m_{K^*} = m_{\bar{K}^*}$.

\section{Numerical results} \label{numericalresults}

In this section we shown the theoretical numerical results obtained for the scattering amplitude square of the three-body $\eta K^*\bar{K}^*$, $\pi K^*\bar{K}^*$, and $KK^*\bar{K}^*$ systems with total isospin $I=0$, $1$, and $1/2$, respectively. We evaluate the three body scattering amplitude $T$ and associate the peaks in the modulus squared $|T|^2$ to resonances. In addition, the masses of the particles considered in this work are shown in Table~\ref{tab:data}. These values are taken from the review of particle physics~\cite{ParticleDataGroup:2020ssz}. For the mass of $f_0(1710)$ state, both "evaluation mass" (eva.) and "average mass" (ave.) are shown.

\begin{table}[htbp]
\centering
\renewcommand\arraystretch{1.5}
\tabcolsep=0.15cm
\caption{Particle masses (in MeV) used in this work.}
    \begin{tabular}{|c|c|c|c|c|c|}
        \hline
         $m_K$&$m_{\pi}$&$m_{\eta}$&$m_{\eta'}$&$\Gamma_{K^*}$&$M_{f_0(1710)}$ (eva.)\\
         \hline
         495.6455&138.04&547.862&957.78&49.1&1704\\
         \hline
         $m_{K^*/\bar{K}^*}$&$m_{\rho}$&$m_{\phi}$&$m_{\omega}$&$\Gamma_{\rho}$&$M_{f_1(1710)}$ (ave.)\\
         \hline
         893.1&775.26&1019.461&782.66&149.1&1733\\
         \hline
    \end{tabular}
    \label{tab:data}
\end{table}

\subsection{ Three-body $\eta K^* \bar{K}^*$ system}

\begin{figure}[htbp]
\centering
\includegraphics[scale=0.35]{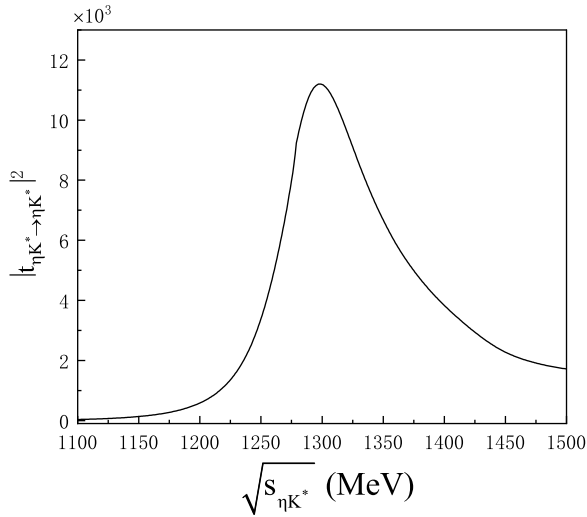}
\caption{Modulus squared of $t_{\eta K^*\rightarrow \eta K^*}$ as a function of the invariant mass of $\eta K^*$ system}
\label{fig:1270}
\end{figure}

In order to obtain two-body system scattering amplitude $t_1$ and $t_2$, we need the scattering amplitude of $\eta K^*$ and $\eta \bar{K}^*$, which were studied in Ref.~\cite{Roca:2005nm,Geng:2006yb}. In this work, we have considered also the affect of $\eta'$ meson as done in Refs.~\cite{Guo:2005wp,Sun:2022cxf}. Within the model parameters as used in Refs.~\cite{Roca:2005nm,Geng:2006yb}: $\mu = 900$ MeV and $\alpha_\mu = -1.85$, one can easily obtain the modulus squared of $t_{\eta K^* \to \eta K^*}$, which is shown in Fig.~\ref{fig:1270}. One can see that there is a clear peak for the $K_1(1270)$ state~\cite{Geng:2006yb}, and the interaction of $\eta K^*$ is strong. Note that the width of these vector mesons are taken into account in the calculations for the loop functions~\cite{Geng:2008gx}, and the affects of those channels including the $\eta'$ meson are very small and could be safely neglected.

On the other hand, one can also get the transition amplitude $t_{\eta \bar{K}^* \to \eta \bar{K}^*}$. Since the $K^*\bar{K}^*$, $\eta K^*$ and $\eta \bar{K}^*$ interactions are attractive and strong enough to generate bound states, it is natural to expect the
existence of multi-hadron states  composed of $K^*\bar{K}^*$ and $\eta$. 

\begin{figure}[htbp]
    \centering
    \includegraphics[width=9cm]{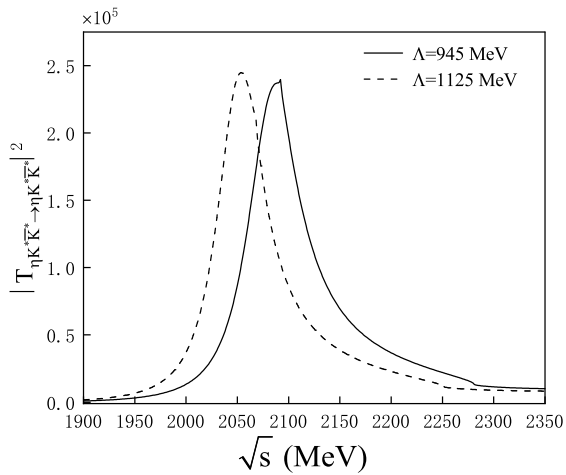}
    \caption{The modulus squared of scattering amplitude for $\eta K^* \bar{K}^*$ three-body system with $\Lambda=945$ MeV (solid line) and $\Lambda=1125$ MeV (dash line).}
    \label{fig:T}
\end{figure}

Then, we obtain the total three body scattering amplitude $T_{\eta(K^*\bar{K}^*)_{f_0(1710)} \to \eta(K^*\bar{K}^*)_{f_0(1710)}}$. In Fig.~\ref{fig:T} we show the modulus squared $|T_{\eta(K^*\bar{K}^*)_{f_0(1710)} \to \eta(K^*\bar{K}^*)_{f_0(1710)}}|^2$ for the $\eta(K^*\bar{K}^*)_{f_0(1710)} \to \eta(K^*\bar{K}^*)_{f_0(1710)}$ scattering amplitude. The picture shows a clear peak around $2100$ MeV with cutoff parameter $\Lambda = 945$ MeV. If one take $\Lambda = 1125$ MeV, the peak moves to $2070$ MeV. This peak with quantum numbers $I^G(J^{PC}) = 0^+(0^{-+})$ could be interpreted as $\eta(2100)$ that have mass $2050_{-24-26}^{+30+75}$ MeV and width $250_{-30-164}^{+36+181}$ MeV. In addition, it is found that the peak is not sensitive to the cutoff parameter $\Lambda$. 

On the experimental side, the $\eta(2100)$ meson was observed firstly by DM2 collaboration in 1988~\cite{DM2:1988esg}. Then it is also observed by BESIII collaboration in a partial wave analysis of the $J/\Phi\rightarrow \gamma \phi \phi$ decay with 22 $\sigma$ significance~\cite{BESIII:2016qzq}. However, this state is not listed in the summary meson tables of the RPP~\cite{ParticleDataGroup:2020ssz}, which indicates more analysis about it are needed. Besides, in Ref.~\cite{Wang:2017iai}, the $\eta(2100)$ meson is interpreted as a candidate for $\eta(4S)$ of the fourth pseudoscalar meson nonet.

\subsection{Three-body $\pi K^* \bar{K}^*$ system}

In the case of $\pi K^* \bar{K}^*$ system, we need the two-body $\pi K^* \to \pi K^*$ scattering amplitude as inputs. Within the formula and theoretical parameters of Ref.~\cite{Roca:2005nm}, one can easily get the two-body scattering amplitude $t_{\pi K^* \to \pi K^*}$ in coupled channels. Note that we use the cutoff regularization scheme for the loop functions of the intermediate vector-pseduoscalar mesons, and the widths of the vector mesons are also considered. Furthermore, we take cutoff parameter $\Lambda_{\pi K^*}=1000$ MeV to regularize the loop function containing a vector and a pseudoscalar meson. 

\begin{figure}[htbp]
    \includegraphics[scale=0.35]{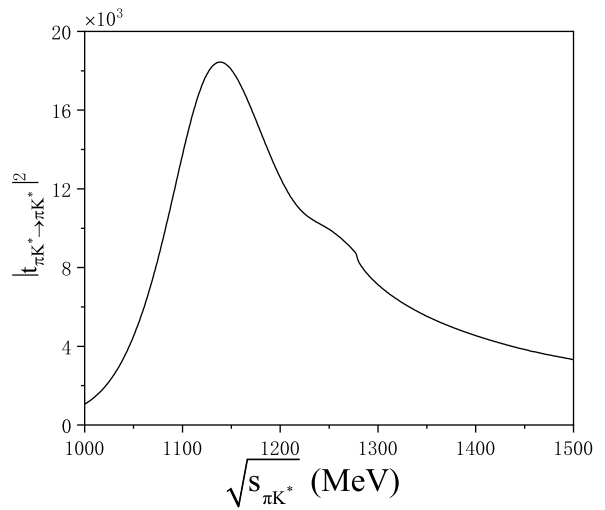}
     \caption{The modulus squared of scattering amplitude of $t_{\pi K^* \rightarrow \pi K^*}$ in isospin $I=1/2$ sector as the function of $\pi K^*$ system.}
    \label{fig:pi-cutoff}
\end{figure}

The modulus squared of $t_{\pi K^* \to \pi K^*}$ in isospin $I=1/2$ is shown in Fig.~\ref{fig:pi-cutoff}, from where one can see that there is a clear peak around $1150$ MeV. As discussed in Ref.~\cite{Roca:2005nm}, it could be the lower pole of the two $K_1(1270)$ states.

\begin{figure}[htbp]
    \includegraphics[scale=0.35]{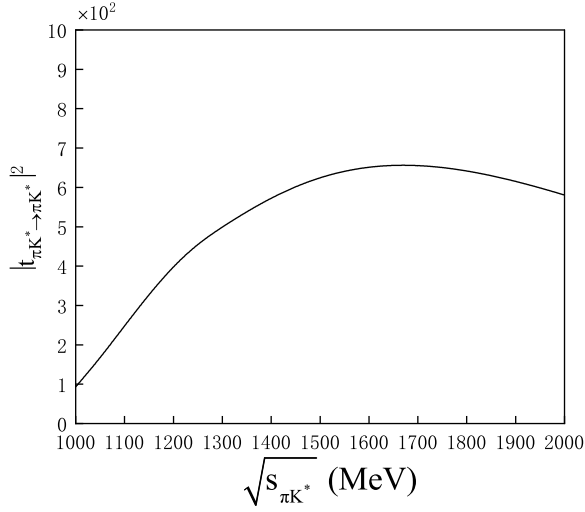}
     \caption{The modulus squared of scattering amplitude of $t_{\pi K^* \rightarrow \pi K^*}$ in isospin $I=3/2$ sector as the function of $\pi K^*$ system.}
    \label{fig:pik-cutoff}
\end{figure}

The interaction of $\pi K^*$ in isospin $I=3/2$ is much weaker then the one with $I=1/2$. The numerical results of $t_{\pi K^* \to \pi K^*}$ with $I=3/2$ are shown in Fig.~\ref{fig:pik-cutoff}. On can see that the strength of $t^{I=3/2}_{\pi K^* \to \pi K^*}$ is much smaller than the one of $t^{I=1/2}_{\pi K^* \to \pi K^*}$ and there is no clear bump structure of the$|t^{I=3/2}_{\pi K^* \to \pi K^*}|^2$ in a wide energy region.

\begin{figure}[htbp]
        \centering
        \includegraphics[scale=0.36]{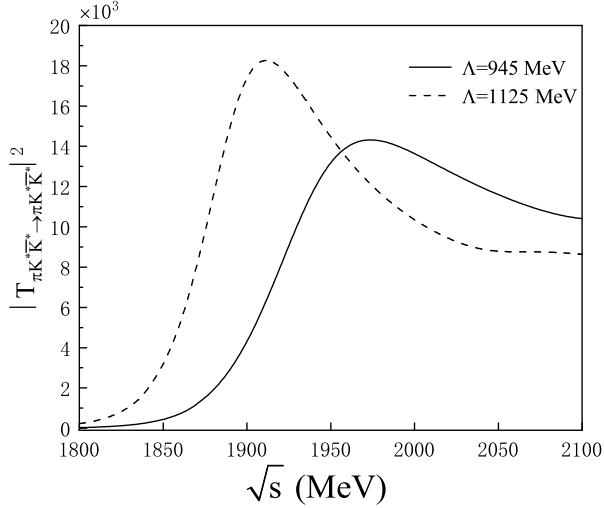}
        \caption{The modulus squared of scattering amplitude for $\pi K^* \bar{K}^*$ three-body system with $\Lambda=945$ MeV (solid line) and $\Lambda=1125 $ MeV (dash line)}
        \label{fig:T-pi}
    \end{figure}

In Fig.~\ref{fig:T-pi}, we show the obtained modulus squared of scattering amplitude $|T_{\pi (K^*\bar{K}^*)_{f_0(1710)} \to \pi (K^*\bar{K}^*)_{f_0(1710)}}|^2$ of the three-body $\pi K^*\bar{K}^*$ system. It is found that there is a wide bump structure around $1900$-$2000$ MeV, which may be associated to the $\pi(2070)$ state, which was not quoted in the RPP~\cite{ParticleDataGroup:2020ssz}. Around that energy region, the $\pi(2070)$ state was needed in a combined partial wave analysis of $\bar{p} p$ annihilation channels in Ref.~\cite{Anisovich:2001pn}, and its mass and width are about $2070 \pm 35$ MeV and $310^{+100}_{-50}$ MeV, respectively. The obtained width of $\pi(2070)$ state is rather wide. Thus the mass of $\pi(2070)$ state is needed to be precisely measured by further experiment since its large width will lead to difficulties for the measurements.

\subsection{Three-body $K K^* \bar{K}^*$ system}

The interaction between $K$ and $\bar{K}^*$ is strong, and the $f_1(1285)$ state is dynamically generated from the $K\bar{K}^*$ interaction in $S$-wave and isospin $I=0$ sector~\cite{Roca:2005nm}. While in the $I=1$ sector, the $a_1(1260)$ and $b_1(1235)$ can be also obtained~\cite{Roca:2005nm}. Within the same parameters as used in Ref.~\cite{Roca:2005nm}, the obtained modulus squared of $t_{K\bar{K}^*\to K\bar{K}^*}$ in $I=0$ and $I=1$ are shown in Figs.~\ref{fig:0,0} and \ref{fig:0,1}, respectively. From these figures, one can see that there are clearly the peaks for the $f_1(1285)$ and $a_1(1260)/b_1(1235)$ states, and the former peak is much narrower, yet, the later one is wide and the peak is superposition of $a_1(1260)$ and $b_1(1235)$, since in the calculations, we don not distinguish their $G$-parity. Furthermore, the interaction of $K\bar{K}^*$ in $I=1$ sector is much weaker than that in the $I=0$ sector.
   
\begin{figure}[htbp]
       \centering
       \includegraphics[width=9cm]{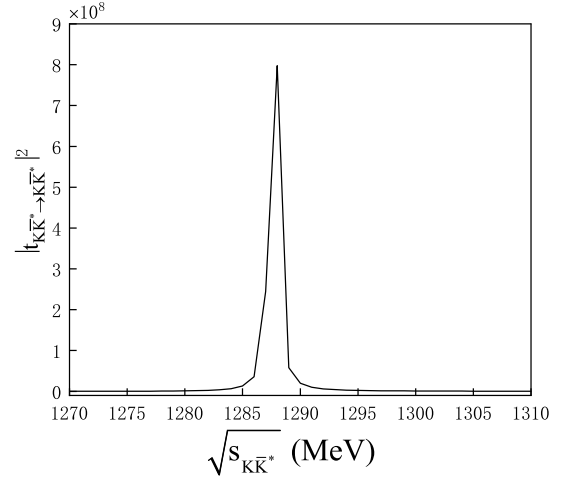}
       \caption{Modulus squared of $t_{K\bar{K}^*\rightarrow K\bar{K}^*}$ in $I=0$ as a function of the invariant mass of $K\bar{K}^*$ system }
       \label{fig:0,0}
\end{figure}
   
\begin{figure}[htbp]
       \centering
       \includegraphics[width=9cm]{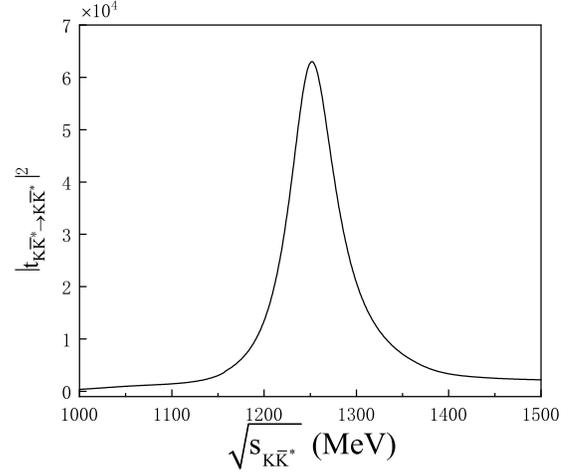}
       \caption{Modulus squared of $t_{K\bar{K}^*\rightarrow K\bar{K}^*}$ in $I=1$ as a function of the invariant mass of $K\bar{K}^*$ system }
       \label{fig:0,1}
\end{figure}
   
On the other hand, the interaction between $K$ and $K^*$ is even weaker, and the two body scattering amplitude of $t_{K K^* \to KK^*}$ in $I=0$ is zero. In the $I=1$ sector, the scattering amplitude squared of $t_{K K^* \to KK^*}$ is shown in Fig.~\ref{fig:2,1}. The strength of $|t_{K K^* \to KK^*}|^2$ is much smaller than those shown in Figs.~\ref{fig:0,0} and \ref{fig:0,1} for the case of $K \bar{K}^*$ scattering.
 
\begin{figure}[htbp]
       \centering
       \includegraphics[scale=0.36]{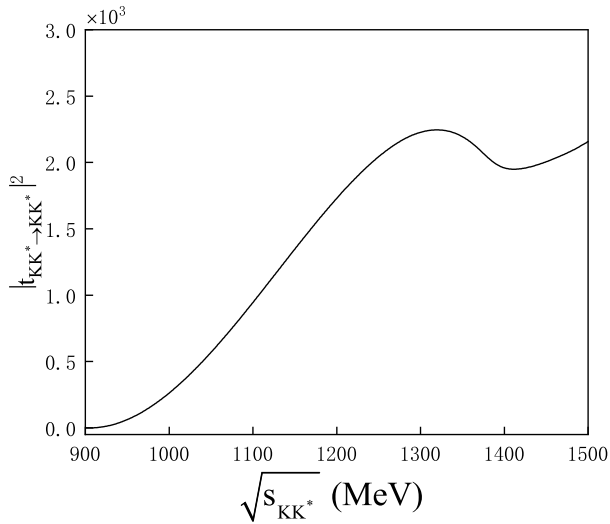}
       \caption{Modulus squared of $t_{K K^*\rightarrow K K^*}$ in $I=1$ as a function of the invariant mass of $K K^*$ system }
       \label{fig:2,1}
   \end{figure}

\begin{figure}[htbp]
    \centering
    \includegraphics[scale=0.36]{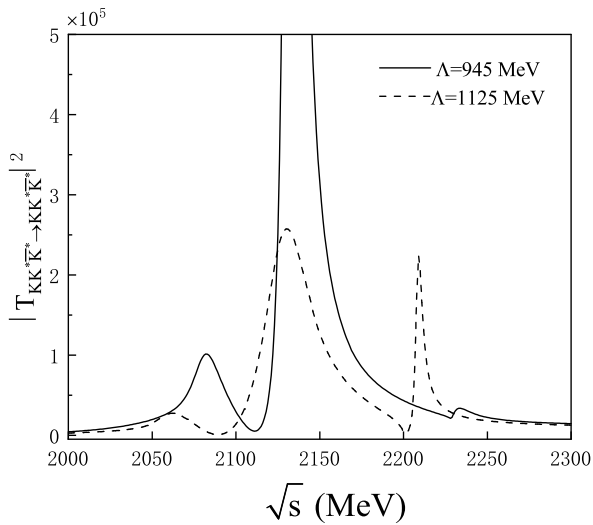}
    \caption{The modulus squared of scattering amplitude for $KK^*\bar{K}^*$ three-body system with $\Lambda=945$ MeV (solid line) and $\Lambda=1125$ MeV (dash line).}
    \label{fig:T-k}
\end{figure}

In Fig.~\ref{fig:T-k}, the modulus squared of scattering amplitude of three-body $K K^*\bar{K}^*$ system is shown, where there are three peaks. The one around $2130$ MeV is not sensitive to the cut off parameter, while the other two are much dependent on the value of cut off parameter $\Lambda$. The lowest one is located at about 2072 MeV, and the highest one is about 2219 MeV. 

On the experimental side, there is a meson with strangeness around 2062 MeV and named $X(2075)$. It was observed by BESII collaboration in the $p\bar{\Lambda}$ invariant mass spectrum of the decay $J/\psi \rightarrow K^- p \bar{\Lambda}$~\cite{BES:2004fgd}. Its Breit-Wigner mass and width are $2075 \pm 12 ({\rm stat.}) \pm 5 ({\rm syst.})$ MeV and $90 \pm 35 ({\rm stat.}) \pm 9 ({\rm syst.})$ MeV, respectively. The spin of $X(2075)$ state could be 0 or 1. Note that, as pointed in Ref.~\cite{BES:2004fgd}, its interpretation as a conventional $K^*$ meson would be disfavored.

In addition, a similar near-threshold enhancement in the $p \bar{\Lambda}$ system was also observed in $B^+ \rightarrow p \bar{\Lambda} D^0$ by Belle collaboration~\cite{Belle:2011cxw}. In fact, one can notice that there are a large blank space between 2000-3000 MeV in RPP for those strangeness states~\cite{ParticleDataGroup:2020ssz}. We look forward to further experiment can give us more information about those mesons with a strange quark in this energy region. 

\section{Summary}

In the framework of fixed-center approximation, we have performed a calculation for the three-body $\eta (K^*\bar{K}^*)_{f_0(1710)}$, $\pi (K^*\bar{K}^*)_{f_0(1710)}$, and $K (K^*\bar{K}^*)_{f_0(1710)}$ systems treating $f_0(1710)$ as a $K^*\bar{K}^*$ bound state as found in previous studies of the vector-vector interactions. The total three-body scattering amplitude can be obtained in terms of the two-body interactions which are taken from the chiral unitary approach. We associate the peaks in the modulus squared of the three-body scattering amplitude to hadronic states. 

Our theoretical results for the obtained mass of $f_0(1710)$ state and these peak positions of squared of three-body scattering amplitudes with different values of the cutoff parameter $\Lambda$ are listed in Table~\ref{tab:cutoff-peak}. The peak positions are associated to the mass of the three-body state. The state found in the $\eta K^* \bar{K}^*$ system could be interpreted as $\eta(2100)$ meson. For the $\pi K^*\bar{K}^*$ system, we find a bump structure between 1900 and 2000 MeV. While for the $K K^*\bar{K}^*$ system, there are three peaks appeared. One of them is not sensitive to the cutoff parameter, and its mass is about 2130 MeV. The third one is disappeared if the cutoff parameter less than $855$ MeV. It is expected the theoretical calculations here could be tested by future experiments.

\begin{table}[htbp]
    \centering
    \renewcommand\arraystretch{1.5}
    \tabcolsep=0.2cm
     \caption{Peak positions of squared of three-body scattering amplitude corresponding to different value of cut off parameter $\Lambda$. Unit in MeV.}
    \label{tab:cutoff-peak}
    \begin{tabular}{|c|c|c|c|c|c|c|}
    \hline
    \multicolumn{2}{|c|}{$\Lambda$} &855&945&1035&1125&1215 \\
    \hline
    \multicolumn{2}{|c|}{$M_{f_0(1710)}$}&1747&1733&1718&1704&1690\\
    \hline
    \multicolumn{2}{|c|}{$M_{\eta K^*\bar{K}^*}$}&2104&2092&2070&2054&2038\\
    \hline
    \multicolumn{2}{|c|}{$M_{\pi K^*\bar{K}^*}$}&1981&1963&1948&1929&1913\\  
    \hline
    \multirow{3}{*}{$M_{K K^*\bar{K}^*}$}
    & state 1 &2092&2083&2072&2062&2053\\
   \cline{2-7}
    & state 2 &2140&2134&2130&2130&2134\\
    \cline{2-7}
    & state 3 & $--$ & 2233& 2219& 2209& 2211\\
    \hline
    \end{tabular}
\end{table}

\begin{acknowledgments}
We thank Xu Zhang for useful discussions. This work is partly supported by the National Natural Science Foundation of China under Grant Nos. 12075288, 11735003, and 11961141012. It is also supported by the Youth Innovation Promotion Association CAS.
\end{acknowledgments}

\bibliographystyle{apsrev4-1.bst}
\bibliography{references.bib}

\begin{thebibliography}{64}%
\makeatletter
\providecommand \@ifxundefined [1]{%
 \@ifx{#1\undefined}
}%
\providecommand \@ifnum [1]{%
 \ifnum #1\expandafter \@firstoftwo
 \else \expandafter \@secondoftwo
 \fi
}%
\providecommand \@ifx [1]{%
 \ifx #1\expandafter \@firstoftwo
 \else \expandafter \@secondoftwo
 \fi
}%
\providecommand \natexlab [1]{#1}%
\providecommand \enquote  [1]{``#1''}%
\providecommand \bibnamefont  [1]{#1}%
\providecommand \bibfnamefont [1]{#1}%
\providecommand \citenamefont [1]{#1}%
\providecommand \href@noop [0]{\@secondoftwo}%
\providecommand \href [0]{\begingroup \@sanitize@url \@href}%
\providecommand \@href[1]{\@@startlink{#1}\@@href}%
\providecommand \@@href[1]{\endgroup#1\@@endlink}%
\providecommand \@sanitize@url [0]{\catcode `\\12\catcode `\$12\catcode
  `\&12\catcode `\#12\catcode `\^12\catcode `\_12\catcode `\%12\relax}%
\providecommand \@@startlink[1]{}%
\providecommand \@@endlink[0]{}%
\providecommand \url  [0]{\begingroup\@sanitize@url \@url }%
\providecommand \@url [1]{\endgroup\@href {#1}{\urlprefix }}%
\providecommand \urlprefix  [0]{URL }%
\providecommand \Eprint [0]{\href }%
\providecommand \doibase [0]{http://dx.doi.org/}%
\providecommand \selectlanguage [0]{\@gobble}%
\providecommand \bibinfo  [0]{\@secondoftwo}%
\providecommand \bibfield  [0]{\@secondoftwo}%
\providecommand \translation [1]{[#1]}%
\providecommand \BibitemOpen [0]{}%
\providecommand \bibitemStop [0]{}%
\providecommand \bibitemNoStop [0]{.\EOS\space}%
\providecommand \EOS [0]{\spacefactor3000\relax}%
\providecommand \BibitemShut  [1]{\csname bibitem#1\endcsname}%
\let\auto@bib@innerbib\@empty
\bibitem [{\citenamefont {Oset}\ \emph {et~al.}(2016)\citenamefont {Oset} \emph
  {et~al.}}]{Oset:2016lyh}%
  \BibitemOpen
  \bibfield  {author} {\bibinfo {author} {\bibfnamefont {E.}~\bibnamefont
  {Oset}} \emph {et~al.},\ }\href {\doibase 10.1142/S0218301316300010}
  {\bibfield  {journal} {\bibinfo  {journal} {Int. J. Mod. Phys. E}\ }\textbf
  {\bibinfo {volume} {25}},\ \bibinfo {pages} {1630001} (\bibinfo {year}
  {2016})},\ \Eprint {http://arxiv.org/abs/1601.03972} {arXiv:1601.03972
  [hep-ph]} \BibitemShut {NoStop}%
\bibitem [{\citenamefont {Guo}\ \emph {et~al.}(2018)\citenamefont {Guo},
  \citenamefont {Hanhart}, \citenamefont {Mei\ss{}ner}, \citenamefont {Wang},
  \citenamefont {Zhao},\ and\ \citenamefont {Zou}}]{Guo:2017jvc}%
  \BibitemOpen
  \bibfield  {author} {\bibinfo {author} {\bibfnamefont {F.-K.}\ \bibnamefont
  {Guo}}, \bibinfo {author} {\bibfnamefont {C.}~\bibnamefont {Hanhart}},
  \bibinfo {author} {\bibfnamefont {U.-G.}\ \bibnamefont {Mei\ss{}ner}},
  \bibinfo {author} {\bibfnamefont {Q.}~\bibnamefont {Wang}}, \bibinfo {author}
  {\bibfnamefont {Q.}~\bibnamefont {Zhao}}, \ and\ \bibinfo {author}
  {\bibfnamefont {B.-S.}\ \bibnamefont {Zou}},\ }\href {\doibase
  10.1103/RevModPhys.90.015004} {\bibfield  {journal} {\bibinfo  {journal}
  {Rev. Mod. Phys.}\ }\textbf {\bibinfo {volume} {90}},\ \bibinfo {pages}
  {015004} (\bibinfo {year} {2018})},\ \bibinfo {note} {[Erratum: Rev.Mod.Phys.
  94, 029901 (2022)]},\ \Eprint {http://arxiv.org/abs/1705.00141}
  {arXiv:1705.00141} \BibitemShut {NoStop}%
\bibitem [{\citenamefont {Dong}\ \emph
  {et~al.}(2021{\natexlab{a}})\citenamefont {Dong}, \citenamefont {Guo},\ and\
  \citenamefont {Zou}}]{Dong:2021juy}%
  \BibitemOpen
  \bibfield  {author} {\bibinfo {author} {\bibfnamefont {X.-K.}\ \bibnamefont
  {Dong}}, \bibinfo {author} {\bibfnamefont {F.-K.}\ \bibnamefont {Guo}}, \
  and\ \bibinfo {author} {\bibfnamefont {B.-S.}\ \bibnamefont {Zou}},\ }\href
  {\doibase 10.13725/j.cnki.pip.2021.02.001} {\bibfield  {journal} {\bibinfo
  {journal} {Progr. Phys.}\ }\textbf {\bibinfo {volume} {41}},\ \bibinfo
  {pages} {65} (\bibinfo {year} {2021}{\natexlab{a}})},\ \Eprint
  {http://arxiv.org/abs/2101.01021} {arXiv:2101.01021 [hep-ph]} \BibitemShut
  {NoStop}%
\bibitem [{\citenamefont {Dong}\ \emph
  {et~al.}(2021{\natexlab{b}})\citenamefont {Dong}, \citenamefont {Guo},\ and\
  \citenamefont {Zou}}]{Dong:2021bvy}%
  \BibitemOpen
  \bibfield  {author} {\bibinfo {author} {\bibfnamefont {X.-K.}\ \bibnamefont
  {Dong}}, \bibinfo {author} {\bibfnamefont {F.-K.}\ \bibnamefont {Guo}}, \
  and\ \bibinfo {author} {\bibfnamefont {B.-S.}\ \bibnamefont {Zou}},\ }\href
  {\doibase 10.1088/1572-9494/ac27a2} {\bibfield  {journal} {\bibinfo
  {journal} {Commun. Theor. Phys.}\ }\textbf {\bibinfo {volume} {73}},\
  \bibinfo {pages} {125201} (\bibinfo {year} {2021}{\natexlab{b}})},\ \Eprint
  {http://arxiv.org/abs/2108.02673} {arXiv:2108.02673 [hep-ph]} \BibitemShut
  {NoStop}%
\bibitem [{\citenamefont {Ablikim}\ \emph
  {et~al.}(2022{\natexlab{a}})\citenamefont {Ablikim} \emph
  {et~al.}}]{BESIII:2022iwi}%
  \BibitemOpen
  \bibfield  {author} {\bibinfo {author} {\bibfnamefont {M.}~\bibnamefont
  {Ablikim}} \emph {et~al.} (\bibinfo {collaboration} {BESIII}),\ }\href
  {\doibase 10.1103/PhysRevD.106.072012} {\bibfield  {journal} {\bibinfo
  {journal} {Phys. Rev. D}\ }\textbf {\bibinfo {volume} {106}},\ \bibinfo
  {pages} {072012} (\bibinfo {year} {2022}{\natexlab{a}})},\ \Eprint
  {http://arxiv.org/abs/2202.00623} {arXiv:2202.00623 [hep-ex]} \BibitemShut
  {NoStop}%
\bibitem [{\citenamefont {Ablikim}\ \emph
  {et~al.}(2022{\natexlab{b}})\citenamefont {Ablikim} \emph
  {et~al.}}]{BESIII:2022riz}%
  \BibitemOpen
  \bibfield  {author} {\bibinfo {author} {\bibfnamefont {M.}~\bibnamefont
  {Ablikim}} \emph {et~al.} (\bibinfo {collaboration} {BESIII}),\ }\href
  {\doibase 10.1103/PhysRevLett.129.192002} {\bibfield  {journal} {\bibinfo
  {journal} {Phys. Rev. Lett.}\ }\textbf {\bibinfo {volume} {129}},\ \bibinfo
  {pages} {192002} (\bibinfo {year} {2022}{\natexlab{b}})},\ \Eprint
  {http://arxiv.org/abs/2202.00621} {arXiv:2202.00621 [hep-ex]} \BibitemShut
  {NoStop}%
\bibitem [{\citenamefont {Yang}\ and\ \citenamefont
  {Huang}(2022)}]{Yang:2022lwq}%
  \BibitemOpen
  \bibfield  {author} {\bibinfo {author} {\bibfnamefont {F.}~\bibnamefont
  {Yang}}\ and\ \bibinfo {author} {\bibfnamefont {Y.}~\bibnamefont {Huang}},\
  }\href@noop {} {\  (\bibinfo {year} {2022})},\ \Eprint
  {http://arxiv.org/abs/2203.06934} {arXiv:2203.06934 [hep-ph]} \BibitemShut
  {NoStop}%
\bibitem [{\citenamefont {Dong}\ \emph {et~al.}(2022)\citenamefont {Dong},
  \citenamefont {Lin},\ and\ \citenamefont {Zou}}]{Dong:2022cuw}%
  \BibitemOpen
  \bibfield  {author} {\bibinfo {author} {\bibfnamefont {X.-K.}\ \bibnamefont
  {Dong}}, \bibinfo {author} {\bibfnamefont {Y.-H.}\ \bibnamefont {Lin}}, \
  and\ \bibinfo {author} {\bibfnamefont {B.-S.}\ \bibnamefont {Zou}},\ }\href
  {\doibase 10.1007/s11433-022-1887-5} {\bibfield  {journal} {\bibinfo
  {journal} {Sci. China Phys. Mech. Astron.}\ }\textbf {\bibinfo {volume}
  {65}},\ \bibinfo {pages} {261011} (\bibinfo {year} {2022})},\ \Eprint
  {http://arxiv.org/abs/2202.00863} {arXiv:2202.00863 [hep-ph]} \BibitemShut
  {NoStop}%
\bibitem [{\citenamefont {Chen}\ \emph {et~al.}(2022)\citenamefont {Chen},
  \citenamefont {Su},\ and\ \citenamefont {Zhu}}]{Chen:2022qpd}%
  \BibitemOpen
  \bibfield  {author} {\bibinfo {author} {\bibfnamefont {H.-X.}\ \bibnamefont
  {Chen}}, \bibinfo {author} {\bibfnamefont {N.}~\bibnamefont {Su}}, \ and\
  \bibinfo {author} {\bibfnamefont {S.-L.}\ \bibnamefont {Zhu}},\ }\href
  {\doibase 10.1088/0256-307X/39/5/051201} {\bibfield  {journal} {\bibinfo
  {journal} {Chin. Phys. Lett.}\ }\textbf {\bibinfo {volume} {39}},\ \bibinfo
  {pages} {051201} (\bibinfo {year} {2022})},\ \Eprint
  {http://arxiv.org/abs/2202.04918} {arXiv:2202.04918} \BibitemShut {NoStop}%
\bibitem [{\citenamefont {Zhang}\ \emph {et~al.}(2017)\citenamefont {Zhang},
  \citenamefont {Xie},\ and\ \citenamefont {Chen}}]{Zhang:2016bmy}%
  \BibitemOpen
  \bibfield  {author} {\bibinfo {author} {\bibfnamefont {X.}~\bibnamefont
  {Zhang}}, \bibinfo {author} {\bibfnamefont {J.-J.}\ \bibnamefont {Xie}}, \
  and\ \bibinfo {author} {\bibfnamefont {X.}~\bibnamefont {Chen}},\ }\href
  {\doibase 10.1103/PhysRevD.95.056014} {\bibfield  {journal} {\bibinfo
  {journal} {Phys. Rev. D}\ }\textbf {\bibinfo {volume} {95}},\ \bibinfo
  {pages} {056014} (\bibinfo {year} {2017})},\ \Eprint
  {http://arxiv.org/abs/1612.02613} {arXiv:1612.02613 [hep-ph]} \BibitemShut
  {NoStop}%
\bibitem [{\citenamefont {Lutz}\ and\ \citenamefont
  {Kolomeitsev}(2004)}]{Lutz:2003fm}%
  \BibitemOpen
  \bibfield  {author} {\bibinfo {author} {\bibfnamefont {M.}~\bibnamefont
  {Lutz}}\ and\ \bibinfo {author} {\bibfnamefont {E.}~\bibnamefont
  {Kolomeitsev}},\ }\href {\doibase 10.1016/j.nuclphysa.2003.11.009} {\bibfield
   {journal} {\bibinfo  {journal} {Nuclear Physics A}\ }\textbf {\bibinfo
  {volume} {730}},\ \bibinfo {pages} {392} (\bibinfo {year}
  {2004})}\BibitemShut {NoStop}%
\bibitem [{\citenamefont {Roca}\ \emph {et~al.}(2005)\citenamefont {Roca},
  \citenamefont {Oset},\ and\ \citenamefont {Singh}}]{Roca:2005nm}%
  \BibitemOpen
  \bibfield  {author} {\bibinfo {author} {\bibfnamefont {L.}~\bibnamefont
  {Roca}}, \bibinfo {author} {\bibfnamefont {E.}~\bibnamefont {Oset}}, \ and\
  \bibinfo {author} {\bibfnamefont {J.}~\bibnamefont {Singh}},\ }\href
  {\doibase 10.1103/PhysRevD.72.014002} {\bibfield  {journal} {\bibinfo
  {journal} {Phys. Rev. D}\ }\textbf {\bibinfo {volume} {72}},\ \bibinfo
  {pages} {014002} (\bibinfo {year} {2005})},\ \Eprint
  {http://arxiv.org/abs/hep-ph/0503273} {arXiv:hep-ph/0503273} \BibitemShut
  {NoStop}%
\bibitem [{\citenamefont {Zhou}\ \emph {et~al.}(2014)\citenamefont {Zhou},
  \citenamefont {Ren}, \citenamefont {Chen},\ and\ \citenamefont
  {Geng}}]{Zhou:2014ila}%
  \BibitemOpen
  \bibfield  {author} {\bibinfo {author} {\bibfnamefont {Y.}~\bibnamefont
  {Zhou}}, \bibinfo {author} {\bibfnamefont {X.-L.}\ \bibnamefont {Ren}},
  \bibinfo {author} {\bibfnamefont {H.-X.}\ \bibnamefont {Chen}}, \ and\
  \bibinfo {author} {\bibfnamefont {L.-S.}\ \bibnamefont {Geng}},\ }\href
  {\doibase 10.1103/PhysRevD.90.014020} {\bibfield  {journal} {\bibinfo
  {journal} {Phys. Rev. D}\ }\textbf {\bibinfo {volume} {90}},\ \bibinfo
  {pages} {014020} (\bibinfo {year} {2014})},\ \Eprint
  {http://arxiv.org/abs/1404.6847} {arXiv:1404.6847 [nucl-th]} \BibitemShut
  {NoStop}%
\bibitem [{\citenamefont {Geng}\ \emph {et~al.}(2015)\citenamefont {Geng},
  \citenamefont {Ren}, \citenamefont {Zhou}, \citenamefont {Chen},\ and\
  \citenamefont {Oset}}]{Geng:2015yta}%
  \BibitemOpen
  \bibfield  {author} {\bibinfo {author} {\bibfnamefont {L.-S.}\ \bibnamefont
  {Geng}}, \bibinfo {author} {\bibfnamefont {X.-L.}\ \bibnamefont {Ren}},
  \bibinfo {author} {\bibfnamefont {Y.}~\bibnamefont {Zhou}}, \bibinfo {author}
  {\bibfnamefont {H.-X.}\ \bibnamefont {Chen}}, \ and\ \bibinfo {author}
  {\bibfnamefont {E.}~\bibnamefont {Oset}},\ }\href {\doibase
  10.1103/PhysRevD.92.014029} {\bibfield  {journal} {\bibinfo  {journal} {Phys.
  Rev. D}\ }\textbf {\bibinfo {volume} {92}},\ \bibinfo {pages} {014029}
  (\bibinfo {year} {2015})},\ \Eprint {http://arxiv.org/abs/1503.06633}
  {arXiv:1503.06633 [hep-ph]} \BibitemShut {NoStop}%
\bibitem [{\citenamefont {Xie}\ \emph {et~al.}(2020)\citenamefont {Xie},
  \citenamefont {Li},\ and\ \citenamefont {Liu}}]{Xie:2019iwz}%
  \BibitemOpen
  \bibfield  {author} {\bibinfo {author} {\bibfnamefont {J.-J.}\ \bibnamefont
  {Xie}}, \bibinfo {author} {\bibfnamefont {G.}~\bibnamefont {Li}}, \ and\
  \bibinfo {author} {\bibfnamefont {X.-H.}\ \bibnamefont {Liu}},\ }\href
  {\doibase 10.1088/1674-1137/abae51} {\bibfield  {journal} {\bibinfo
  {journal} {Chin. Phys. C}\ }\textbf {\bibinfo {volume} {44}},\ \bibinfo
  {pages} {114104} (\bibinfo {year} {2020})},\ \Eprint
  {http://arxiv.org/abs/1907.12202} {arXiv:1907.12202 [hep-ph]} \BibitemShut
  {NoStop}%
\bibitem [{\citenamefont {Zhang}\ and\ \citenamefont
  {Xie}(2020)}]{Zhang:2019ykd}%
  \BibitemOpen
  \bibfield  {author} {\bibinfo {author} {\bibfnamefont {X.}~\bibnamefont
  {Zhang}}\ and\ \bibinfo {author} {\bibfnamefont {J.-J.}\ \bibnamefont
  {Xie}},\ }\href {\doibase 10.1088/1674-1137/44/5/054104} {\bibfield
  {journal} {\bibinfo  {journal} {Chin. Phys. C}\ }\textbf {\bibinfo {volume}
  {44}},\ \bibinfo {pages} {054104} (\bibinfo {year} {2020})},\ \Eprint
  {http://arxiv.org/abs/1906.07340} {arXiv:1906.07340 [nucl-th]} \BibitemShut
  {NoStop}%
\bibitem [{\citenamefont {Martinez~Torres}\ \emph {et~al.}(2020)\citenamefont
  {Martinez~Torres}, \citenamefont {Khemchandani}, \citenamefont {Roca},\ and\
  \citenamefont {Oset}}]{MartinezTorres:2020hus}%
  \BibitemOpen
  \bibfield  {author} {\bibinfo {author} {\bibfnamefont {A.}~\bibnamefont
  {Martinez~Torres}}, \bibinfo {author} {\bibfnamefont {K.~P.}\ \bibnamefont
  {Khemchandani}}, \bibinfo {author} {\bibfnamefont {L.}~\bibnamefont {Roca}},
  \ and\ \bibinfo {author} {\bibfnamefont {E.}~\bibnamefont {Oset}},\ }\href
  {\doibase 10.1007/s00601-020-01568-y} {\bibfield  {journal} {\bibinfo
  {journal} {Few Body Syst.}\ }\textbf {\bibinfo {volume} {61}},\ \bibinfo
  {pages} {35} (\bibinfo {year} {2020})},\ \Eprint
  {http://arxiv.org/abs/2005.14357} {arXiv:2005.14357 [nucl-th]} \BibitemShut
  {NoStop}%
\bibitem [{\citenamefont {Wu}\ \emph {et~al.}(2022)\citenamefont {Wu},
  \citenamefont {Pan}, \citenamefont {Liu},\ and\ \citenamefont
  {Geng}}]{Wu:2022ftm}%
  \BibitemOpen
  \bibfield  {author} {\bibinfo {author} {\bibfnamefont {T.-W.}\ \bibnamefont
  {Wu}}, \bibinfo {author} {\bibfnamefont {Y.-W.}\ \bibnamefont {Pan}},
  \bibinfo {author} {\bibfnamefont {M.-Z.}\ \bibnamefont {Liu}}, \ and\
  \bibinfo {author} {\bibfnamefont {L.-S.}\ \bibnamefont {Geng}},\ }\href
  {\doibase 10.1016/j.scib.2022.08.007} {\bibfield  {journal} {\bibinfo
  {journal} {Sci. Bull.}\ }\textbf {\bibinfo {volume} {67}},\ \bibinfo {pages}
  {1735} (\bibinfo {year} {2022})},\ \Eprint {http://arxiv.org/abs/2208.00882}
  {arXiv:2208.00882 [hep-ph]} \BibitemShut {NoStop}%
\bibitem [{\citenamefont {Malabarba}\ \emph {et~al.}(2022)\citenamefont
  {Malabarba}, \citenamefont {Khemchandani},\ and\ \citenamefont
  {Torres}}]{Malabarba:2021taj}%
  \BibitemOpen
  \bibfield  {author} {\bibinfo {author} {\bibfnamefont {B.~B.}\ \bibnamefont
  {Malabarba}}, \bibinfo {author} {\bibfnamefont {K.~P.}\ \bibnamefont
  {Khemchandani}}, \ and\ \bibinfo {author} {\bibfnamefont {A.~M.}\
  \bibnamefont {Torres}},\ }\href {\doibase 10.1140/epja/s10050-022-00681-2}
  {\bibfield  {journal} {\bibinfo  {journal} {Eur. Phys. J. A}\ }\textbf
  {\bibinfo {volume} {58}},\ \bibinfo {pages} {33} (\bibinfo {year} {2022})},\
  \Eprint {http://arxiv.org/abs/2103.09978} {arXiv:2103.09978 [hep-ph]}
  \BibitemShut {NoStop}%
\bibitem [{\citenamefont {Luo}\ \emph {et~al.}(2022{\natexlab{a}})\citenamefont
  {Luo}, \citenamefont {Wu}, \citenamefont {Liu}, \citenamefont {Geng},\ and\
  \citenamefont {Liu}}]{Luo:2021ggs}%
  \BibitemOpen
  \bibfield  {author} {\bibinfo {author} {\bibfnamefont {S.-Q.}\ \bibnamefont
  {Luo}}, \bibinfo {author} {\bibfnamefont {T.-W.}\ \bibnamefont {Wu}},
  \bibinfo {author} {\bibfnamefont {M.-Z.}\ \bibnamefont {Liu}}, \bibinfo
  {author} {\bibfnamefont {L.-S.}\ \bibnamefont {Geng}}, \ and\ \bibinfo
  {author} {\bibfnamefont {X.}~\bibnamefont {Liu}},\ }\href {\doibase
  10.1103/PhysRevD.105.074033} {\bibfield  {journal} {\bibinfo  {journal}
  {Phys. Rev. D}\ }\textbf {\bibinfo {volume} {105}},\ \bibinfo {pages}
  {074033} (\bibinfo {year} {2022}{\natexlab{a}})},\ \Eprint
  {http://arxiv.org/abs/2111.15079} {arXiv:2111.15079 [hep-ph]} \BibitemShut
  {NoStop}%
\bibitem [{\citenamefont {Luo}\ \emph {et~al.}(2022{\natexlab{b}})\citenamefont
  {Luo}, \citenamefont {Geng},\ and\ \citenamefont {Liu}}]{Luo:2022cun}%
  \BibitemOpen
  \bibfield  {author} {\bibinfo {author} {\bibfnamefont {S.-Q.}\ \bibnamefont
  {Luo}}, \bibinfo {author} {\bibfnamefont {L.-S.}\ \bibnamefont {Geng}}, \
  and\ \bibinfo {author} {\bibfnamefont {X.}~\bibnamefont {Liu}},\ }\href
  {\doibase 10.1103/PhysRevD.106.014017} {\bibfield  {journal} {\bibinfo
  {journal} {Phys. Rev. D}\ }\textbf {\bibinfo {volume} {106}},\ \bibinfo
  {pages} {014017} (\bibinfo {year} {2022}{\natexlab{b}})},\ \Eprint
  {http://arxiv.org/abs/2206.04586} {arXiv:2206.04586 [hep-ph]} \BibitemShut
  {NoStop}%
\bibitem [{\citenamefont {Ikeno}\ \emph {et~al.}(2022)\citenamefont {Ikeno},
  \citenamefont {Bayar},\ and\ \citenamefont {Oset}}]{Ikeno:2022jbb}%
  \BibitemOpen
  \bibfield  {author} {\bibinfo {author} {\bibfnamefont {N.}~\bibnamefont
  {Ikeno}}, \bibinfo {author} {\bibfnamefont {M.}~\bibnamefont {Bayar}}, \ and\
  \bibinfo {author} {\bibfnamefont {E.}~\bibnamefont {Oset}},\ }\href@noop {}
  {\  (\bibinfo {year} {2022})},\ \Eprint {http://arxiv.org/abs/2208.03698}
  {arXiv:2208.03698 [hep-ph]} \BibitemShut {NoStop}%
\bibitem [{\citenamefont {Debastiani}\ \emph {et~al.}(2017)\citenamefont
  {Debastiani}, \citenamefont {Dias},\ and\ \citenamefont
  {Oset}}]{Debastiani:2017vhv}%
  \BibitemOpen
  \bibfield  {author} {\bibinfo {author} {\bibfnamefont {V.~R.}\ \bibnamefont
  {Debastiani}}, \bibinfo {author} {\bibfnamefont {J.~M.}\ \bibnamefont
  {Dias}}, \ and\ \bibinfo {author} {\bibfnamefont {E.}~\bibnamefont {Oset}},\
  }\href {\doibase 10.1103/PhysRevD.96.016014} {\bibfield  {journal} {\bibinfo
  {journal} {Phys. Rev. D}\ }\textbf {\bibinfo {volume} {96}},\ \bibinfo
  {pages} {016014} (\bibinfo {year} {2017})},\ \Eprint
  {http://arxiv.org/abs/1705.09257} {arXiv:1705.09257 [hep-ph]} \BibitemShut
  {NoStop}%
\bibitem [{\citenamefont {Wu}\ \emph {et~al.}(2021)\citenamefont {Wu},
  \citenamefont {Liu},\ and\ \citenamefont {Geng}}]{Wu:2020job}%
  \BibitemOpen
  \bibfield  {author} {\bibinfo {author} {\bibfnamefont {T.-W.}\ \bibnamefont
  {Wu}}, \bibinfo {author} {\bibfnamefont {M.-Z.}\ \bibnamefont {Liu}}, \ and\
  \bibinfo {author} {\bibfnamefont {L.-S.}\ \bibnamefont {Geng}},\ }\href
  {\doibase 10.1103/PhysRevD.103.L031501} {\bibfield  {journal} {\bibinfo
  {journal} {Phys. Rev. D}\ }\textbf {\bibinfo {volume} {103}},\ \bibinfo
  {pages} {L031501} (\bibinfo {year} {2021})},\ \Eprint
  {http://arxiv.org/abs/2012.01134} {arXiv:2012.01134 [hep-ph]} \BibitemShut
  {NoStop}%
\bibitem [{\citenamefont {Wei}\ \emph {et~al.}(2022)\citenamefont {Wei},
  \citenamefont {Shen},\ and\ \citenamefont {Xie}}]{Wei:2022jgc}%
  \BibitemOpen
  \bibfield  {author} {\bibinfo {author} {\bibfnamefont {X.}~\bibnamefont
  {Wei}}, \bibinfo {author} {\bibfnamefont {Q.-H.}\ \bibnamefont {Shen}}, \
  and\ \bibinfo {author} {\bibfnamefont {J.-J.}\ \bibnamefont {Xie}},\ }\href
  {\doibase 10.1140/epjc/s10052-022-10675-5} {\bibfield  {journal} {\bibinfo
  {journal} {Eur. Phys. J. C}\ }\textbf {\bibinfo {volume} {82}},\ \bibinfo
  {pages} {718} (\bibinfo {year} {2022})},\ \Eprint
  {http://arxiv.org/abs/2205.12526} {arXiv:2205.12526 [hep-ph]} \BibitemShut
  {NoStop}%
\bibitem [{\citenamefont {Martinez~Torres}\ \emph {et~al.}(2008)\citenamefont
  {Martinez~Torres}, \citenamefont {Khemchandani}, \citenamefont {Geng},
  \citenamefont {Napsuciale},\ and\ \citenamefont
  {Oset}}]{MartinezTorres:2008gy}%
  \BibitemOpen
  \bibfield  {author} {\bibinfo {author} {\bibfnamefont {A.}~\bibnamefont
  {Martinez~Torres}}, \bibinfo {author} {\bibfnamefont {K.~P.}\ \bibnamefont
  {Khemchandani}}, \bibinfo {author} {\bibfnamefont {L.~S.}\ \bibnamefont
  {Geng}}, \bibinfo {author} {\bibfnamefont {M.}~\bibnamefont {Napsuciale}}, \
  and\ \bibinfo {author} {\bibfnamefont {E.}~\bibnamefont {Oset}},\ }\href
  {\doibase 10.1103/PhysRevD.78.074031} {\bibfield  {journal} {\bibinfo
  {journal} {Phys. Rev. D}\ }\textbf {\bibinfo {volume} {78}},\ \bibinfo
  {pages} {074031} (\bibinfo {year} {2008})},\ \Eprint
  {http://arxiv.org/abs/0801.3635} {arXiv:0801.3635 [nucl-th]} \BibitemShut
  {NoStop}%
\bibitem [{\citenamefont {Ren}\ \emph {et~al.}(2018)\citenamefont {Ren},
  \citenamefont {Malabarba}, \citenamefont {Geng}, \citenamefont
  {Khemchandani},\ and\ \citenamefont {Torres}}]{Ren_2018}%
  \BibitemOpen
  \bibfield  {author} {\bibinfo {author} {\bibfnamefont {X.-L.}\ \bibnamefont
  {Ren}}, \bibinfo {author} {\bibfnamefont {B.~B.}\ \bibnamefont {Malabarba}},
  \bibinfo {author} {\bibfnamefont {L.-S.}\ \bibnamefont {Geng}}, \bibinfo
  {author} {\bibfnamefont {K.}~\bibnamefont {Khemchandani}}, \ and\ \bibinfo
  {author} {\bibfnamefont {A.~M.}\ \bibnamefont {Torres}},\ }\href {\doibase
  10.1016/j.physletb.2018.08.034} {\bibfield  {journal} {\bibinfo  {journal}
  {Physics Letters B}\ }\textbf {\bibinfo {volume} {785}},\ \bibinfo {pages}
  {112} (\bibinfo {year} {2018})}\BibitemShut {NoStop}%
\bibitem [{\citenamefont {Dias}\ \emph {et~al.}(2017)\citenamefont {Dias},
  \citenamefont {Debastiani}, \citenamefont {Roca}, \citenamefont {Sakai},\
  and\ \citenamefont {Oset}}]{Dias:2017miz}%
  \BibitemOpen
  \bibfield  {author} {\bibinfo {author} {\bibfnamefont {J.~M.}\ \bibnamefont
  {Dias}}, \bibinfo {author} {\bibfnamefont {V.~R.}\ \bibnamefont
  {Debastiani}}, \bibinfo {author} {\bibfnamefont {L.}~\bibnamefont {Roca}},
  \bibinfo {author} {\bibfnamefont {S.}~\bibnamefont {Sakai}}, \ and\ \bibinfo
  {author} {\bibfnamefont {E.}~\bibnamefont {Oset}},\ }\href {\doibase
  10.1103/PhysRevD.96.094007} {\bibfield  {journal} {\bibinfo  {journal} {Phys.
  Rev. D}\ }\textbf {\bibinfo {volume} {96}},\ \bibinfo {pages} {094007}
  (\bibinfo {year} {2017})},\ \Eprint {http://arxiv.org/abs/1709.01372}
  {arXiv:1709.01372 [hep-ph]} \BibitemShut {NoStop}%
\bibitem [{\citenamefont {Bayar}\ \emph {et~al.}(2011)\citenamefont {Bayar},
  \citenamefont {Yamagata-Sekihara},\ and\ \citenamefont
  {Oset}}]{Bayar:2011qj}%
  \BibitemOpen
  \bibfield  {author} {\bibinfo {author} {\bibfnamefont {M.}~\bibnamefont
  {Bayar}}, \bibinfo {author} {\bibfnamefont {J.}~\bibnamefont
  {Yamagata-Sekihara}}, \ and\ \bibinfo {author} {\bibfnamefont
  {E.}~\bibnamefont {Oset}},\ }\href {\doibase 10.1103/PhysRevC.84.015209}
  {\bibfield  {journal} {\bibinfo  {journal} {Phys. Rev. C}\ }\textbf {\bibinfo
  {volume} {84}},\ \bibinfo {pages} {015209} (\bibinfo {year} {2011})},\
  \Eprint {http://arxiv.org/abs/1102.2854} {arXiv:1102.2854 [hep-ph]}
  \BibitemShut {NoStop}%
\bibitem [{\citenamefont {Xiao}\ \emph {et~al.}(2011)\citenamefont {Xiao},
  \citenamefont {Bayar},\ and\ \citenamefont {Oset}}]{Xiao:2011rc}%
  \BibitemOpen
  \bibfield  {author} {\bibinfo {author} {\bibfnamefont {C.~W.}\ \bibnamefont
  {Xiao}}, \bibinfo {author} {\bibfnamefont {M.}~\bibnamefont {Bayar}}, \ and\
  \bibinfo {author} {\bibfnamefont {E.}~\bibnamefont {Oset}},\ }\href {\doibase
  10.1103/PhysRevD.84.034037} {\bibfield  {journal} {\bibinfo  {journal} {Phys.
  Rev. D}\ }\textbf {\bibinfo {volume} {84}},\ \bibinfo {pages} {034037}
  (\bibinfo {year} {2011})},\ \Eprint {http://arxiv.org/abs/1106.0459}
  {arXiv:1106.0459 [hep-ph]} \BibitemShut {NoStop}%
\bibitem [{\citenamefont {Bayar}\ \emph {et~al.}(2014)\citenamefont {Bayar},
  \citenamefont {Liang}, \citenamefont {Uchino},\ and\ \citenamefont
  {Xiao}}]{Bayar:2013bta}%
  \BibitemOpen
  \bibfield  {author} {\bibinfo {author} {\bibfnamefont {M.}~\bibnamefont
  {Bayar}}, \bibinfo {author} {\bibfnamefont {W.~H.}\ \bibnamefont {Liang}},
  \bibinfo {author} {\bibfnamefont {T.}~\bibnamefont {Uchino}}, \ and\ \bibinfo
  {author} {\bibfnamefont {C.~W.}\ \bibnamefont {Xiao}},\ }\href {\doibase
  10.1140/epja/i2014-14067-0} {\bibfield  {journal} {\bibinfo  {journal} {Eur.
  Phys. J. A}\ }\textbf {\bibinfo {volume} {50}},\ \bibinfo {pages} {67}
  (\bibinfo {year} {2014})},\ \Eprint {http://arxiv.org/abs/1312.2869}
  {arXiv:1312.2869 [hep-ph]} \BibitemShut {NoStop}%
\bibitem [{\citenamefont {Liang}\ \emph {et~al.}(2013)\citenamefont {Liang},
  \citenamefont {Xiao},\ and\ \citenamefont {Oset}}]{Liang:2013yta}%
  \BibitemOpen
  \bibfield  {author} {\bibinfo {author} {\bibfnamefont {W.}~\bibnamefont
  {Liang}}, \bibinfo {author} {\bibfnamefont {C.~W.}\ \bibnamefont {Xiao}}, \
  and\ \bibinfo {author} {\bibfnamefont {E.}~\bibnamefont {Oset}},\ }\href
  {\doibase 10.1103/PhysRevD.88.114024} {\bibfield  {journal} {\bibinfo
  {journal} {Phys. Rev. D}\ }\textbf {\bibinfo {volume} {88}},\ \bibinfo
  {pages} {114024} (\bibinfo {year} {2013})},\ \Eprint
  {http://arxiv.org/abs/1309.7310} {arXiv:1309.7310 [hep-ph]} \BibitemShut
  {NoStop}%
\bibitem [{\citenamefont {Durkaya}\ and\ \citenamefont
  {Bayar}(2015)}]{Durkaya:2015wra}%
  \BibitemOpen
  \bibfield  {author} {\bibinfo {author} {\bibfnamefont {B.}~\bibnamefont
  {Durkaya}}\ and\ \bibinfo {author} {\bibfnamefont {M.}~\bibnamefont
  {Bayar}},\ }\href {\doibase 10.1103/PhysRevD.92.036006} {\bibfield  {journal}
  {\bibinfo  {journal} {Phys. Rev. D}\ }\textbf {\bibinfo {volume} {92}},\
  \bibinfo {pages} {036006} (\bibinfo {year} {2015})}\BibitemShut {NoStop}%
\bibitem [{\citenamefont {Yamagata-Sekihara}\ \emph {et~al.}(2010)\citenamefont
  {Yamagata-Sekihara}, \citenamefont {Roca},\ and\ \citenamefont
  {Oset}}]{Yamagata-Sekihara:2010muv}%
  \BibitemOpen
  \bibfield  {author} {\bibinfo {author} {\bibfnamefont {J.}~\bibnamefont
  {Yamagata-Sekihara}}, \bibinfo {author} {\bibfnamefont {L.}~\bibnamefont
  {Roca}}, \ and\ \bibinfo {author} {\bibfnamefont {E.}~\bibnamefont {Oset}},\
  }\href {\doibase 10.1103/PhysRevD.82.094017} {\bibfield  {journal} {\bibinfo
  {journal} {Phys. Rev. D}\ }\textbf {\bibinfo {volume} {82}},\ \bibinfo
  {pages} {094017} (\bibinfo {year} {2010})},\ \bibinfo {note} {[Erratum:
  Phys.Rev.D 85, 119905 (2012)]},\ \Eprint {http://arxiv.org/abs/1010.0525}
  {arXiv:1010.0525 [hep-ph]} \BibitemShut {NoStop}%
\bibitem [{\citenamefont {Martinez~Torres}\ \emph {et~al.}(2019)\citenamefont
  {Martinez~Torres}, \citenamefont {Khemchandani},\ and\ \citenamefont
  {Geng}}]{MartinezTorres:2018zbl}%
  \BibitemOpen
  \bibfield  {author} {\bibinfo {author} {\bibfnamefont {A.}~\bibnamefont
  {Martinez~Torres}}, \bibinfo {author} {\bibfnamefont {K.~P.}\ \bibnamefont
  {Khemchandani}}, \ and\ \bibinfo {author} {\bibfnamefont {L.-S.}\
  \bibnamefont {Geng}},\ }\href {\doibase 10.1103/PhysRevD.99.076017}
  {\bibfield  {journal} {\bibinfo  {journal} {Phys. Rev. D}\ }\textbf {\bibinfo
  {volume} {99}},\ \bibinfo {pages} {076017} (\bibinfo {year} {2019})},\
  \Eprint {http://arxiv.org/abs/1809.01059} {arXiv:1809.01059 [hep-ph]}
  \BibitemShut {NoStop}%
\bibitem [{\citenamefont {Sanchez~Sanchez}\ \emph {et~al.}(2018)\citenamefont
  {Sanchez~Sanchez}, \citenamefont {Geng}, \citenamefont {Lu}, \citenamefont
  {Hyodo},\ and\ \citenamefont {Valderrama}}]{SanchezSanchez:2017xtl}%
  \BibitemOpen
  \bibfield  {author} {\bibinfo {author} {\bibfnamefont {M.}~\bibnamefont
  {Sanchez~Sanchez}}, \bibinfo {author} {\bibfnamefont {L.-S.}\ \bibnamefont
  {Geng}}, \bibinfo {author} {\bibfnamefont {J.-X.}\ \bibnamefont {Lu}},
  \bibinfo {author} {\bibfnamefont {T.}~\bibnamefont {Hyodo}}, \ and\ \bibinfo
  {author} {\bibfnamefont {M.~P.}\ \bibnamefont {Valderrama}},\ }\href
  {\doibase 10.1103/PhysRevD.98.054001} {\bibfield  {journal} {\bibinfo
  {journal} {Phys. Rev. D}\ }\textbf {\bibinfo {volume} {98}},\ \bibinfo
  {pages} {054001} (\bibinfo {year} {2018})},\ \Eprint
  {http://arxiv.org/abs/1707.03802} {arXiv:1707.03802 [hep-ph]} \BibitemShut
  {NoStop}%
\bibitem [{\citenamefont {Xie}\ \emph {et~al.}(2011{\natexlab{a}})\citenamefont
  {Xie}, \citenamefont {Martinez~Torres},\ and\ \citenamefont
  {Oset}}]{Xie:2010ig}%
  \BibitemOpen
  \bibfield  {author} {\bibinfo {author} {\bibfnamefont {J.-J.}\ \bibnamefont
  {Xie}}, \bibinfo {author} {\bibfnamefont {A.}~\bibnamefont
  {Martinez~Torres}}, \ and\ \bibinfo {author} {\bibfnamefont {E.}~\bibnamefont
  {Oset}},\ }\href {\doibase 10.1103/PhysRevC.83.065207} {\bibfield  {journal}
  {\bibinfo  {journal} {Phys. Rev. C}\ }\textbf {\bibinfo {volume} {83}},\
  \bibinfo {pages} {065207} (\bibinfo {year} {2011}{\natexlab{a}})},\ \Eprint
  {http://arxiv.org/abs/1010.6164} {arXiv:1010.6164 [nucl-th]} \BibitemShut
  {NoStop}%
\bibitem [{\citenamefont {Xie}\ \emph {et~al.}(2011{\natexlab{b}})\citenamefont
  {Xie}, \citenamefont {Martinez~Torres}, \citenamefont {Oset},\ and\
  \citenamefont {Gonzalez}}]{Xie:2011uw}%
  \BibitemOpen
  \bibfield  {author} {\bibinfo {author} {\bibfnamefont {J.-J.}\ \bibnamefont
  {Xie}}, \bibinfo {author} {\bibfnamefont {A.}~\bibnamefont
  {Martinez~Torres}}, \bibinfo {author} {\bibfnamefont {E.}~\bibnamefont
  {Oset}}, \ and\ \bibinfo {author} {\bibfnamefont {P.}~\bibnamefont
  {Gonzalez}},\ }\href {\doibase 10.1103/PhysRevC.83.055204} {\bibfield
  {journal} {\bibinfo  {journal} {Phys. Rev. C}\ }\textbf {\bibinfo {volume}
  {83}},\ \bibinfo {pages} {055204} (\bibinfo {year} {2011}{\natexlab{b}})},\
  \Eprint {http://arxiv.org/abs/1101.1722} {arXiv:1101.1722 [nucl-th]}
  \BibitemShut {NoStop}%
\bibitem [{\citenamefont {Geng}\ and\ \citenamefont
  {Oset}(2009)}]{Geng:2008gx}%
  \BibitemOpen
  \bibfield  {author} {\bibinfo {author} {\bibfnamefont {L.~S.}\ \bibnamefont
  {Geng}}\ and\ \bibinfo {author} {\bibfnamefont {E.}~\bibnamefont {Oset}},\
  }\href {\doibase 10.1103/PhysRevD.79.074009} {\bibfield  {journal} {\bibinfo
  {journal} {Phys. Rev. D}\ }\textbf {\bibinfo {volume} {79}},\ \bibinfo
  {pages} {074009} (\bibinfo {year} {2009})},\ \Eprint
  {http://arxiv.org/abs/0812.1199} {arXiv:0812.1199 [hep-ph]} \BibitemShut
  {NoStop}%
\bibitem [{\citenamefont {Geng}\ \emph {et~al.}(2009)\citenamefont {Geng},
  \citenamefont {Oset}, \citenamefont {Molina},\ and\ \citenamefont
  {Nicmorus}}]{Geng:2009gb}%
  \BibitemOpen
  \bibfield  {author} {\bibinfo {author} {\bibfnamefont {L.~S.}\ \bibnamefont
  {Geng}}, \bibinfo {author} {\bibfnamefont {E.}~\bibnamefont {Oset}}, \bibinfo
  {author} {\bibfnamefont {R.}~\bibnamefont {Molina}}, \ and\ \bibinfo {author}
  {\bibfnamefont {D.}~\bibnamefont {Nicmorus}},\ }\href {\doibase
  10.22323/1.069.0040} {\bibfield  {journal} {\bibinfo  {journal} {PoS}\
  }\textbf {\bibinfo {volume} {EFT09}},\ \bibinfo {pages} {040} (\bibinfo
  {year} {2009})},\ \Eprint {http://arxiv.org/abs/0905.0419} {arXiv:0905.0419
  [hep-ph]} \BibitemShut {NoStop}%
\bibitem [{\citenamefont {Du}\ \emph {et~al.}(2018)\citenamefont {Du},
  \citenamefont {G\"ulmez}, \citenamefont {Guo}, \citenamefont {Mei\ss{}ner},\
  and\ \citenamefont {Wang}}]{Du:2018gyn}%
  \BibitemOpen
  \bibfield  {author} {\bibinfo {author} {\bibfnamefont {M.-L.}\ \bibnamefont
  {Du}}, \bibinfo {author} {\bibfnamefont {D.}~\bibnamefont {G\"ulmez}},
  \bibinfo {author} {\bibfnamefont {F.-K.}\ \bibnamefont {Guo}}, \bibinfo
  {author} {\bibfnamefont {U.-G.}\ \bibnamefont {Mei\ss{}ner}}, \ and\ \bibinfo
  {author} {\bibfnamefont {Q.}~\bibnamefont {Wang}},\ }\href {\doibase
  10.1140/epjc/s10052-018-6475-8} {\bibfield  {journal} {\bibinfo  {journal}
  {Eur. Phys. J. C}\ }\textbf {\bibinfo {volume} {78}},\ \bibinfo {pages} {988}
  (\bibinfo {year} {2018})},\ \Eprint {http://arxiv.org/abs/1808.09664}
  {arXiv:1808.09664 [hep-ph]} \BibitemShut {NoStop}%
\bibitem [{\citenamefont {Garc\'\i{}a-Recio}\ \emph {et~al.}(2013)\citenamefont
  {Garc\'\i{}a-Recio}, \citenamefont {Geng}, \citenamefont {Nieves},
  \citenamefont {Salcedo}, \citenamefont {Wang},\ and\ \citenamefont
  {Xie}}]{Garcia-Recio:2013uva}%
  \BibitemOpen
  \bibfield  {author} {\bibinfo {author} {\bibfnamefont {C.}~\bibnamefont
  {Garc\'\i{}a-Recio}}, \bibinfo {author} {\bibfnamefont {L.~S.}\ \bibnamefont
  {Geng}}, \bibinfo {author} {\bibfnamefont {J.}~\bibnamefont {Nieves}},
  \bibinfo {author} {\bibfnamefont {L.~L.}\ \bibnamefont {Salcedo}}, \bibinfo
  {author} {\bibfnamefont {E.}~\bibnamefont {Wang}}, \ and\ \bibinfo {author}
  {\bibfnamefont {J.-J.}\ \bibnamefont {Xie}},\ }\href {\doibase
  10.1103/PhysRevD.87.096006} {\bibfield  {journal} {\bibinfo  {journal} {Phys.
  Rev. D}\ }\textbf {\bibinfo {volume} {87}},\ \bibinfo {pages} {096006}
  (\bibinfo {year} {2013})},\ \Eprint {http://arxiv.org/abs/1304.1021}
  {arXiv:1304.1021 [hep-ph]} \BibitemShut {NoStop}%
\bibitem [{\citenamefont {Wang}\ and\ \citenamefont
  {Zou}(2021)}]{Wang:2021jub}%
  \BibitemOpen
  \bibfield  {author} {\bibinfo {author} {\bibfnamefont {Z.-L.}\ \bibnamefont
  {Wang}}\ and\ \bibinfo {author} {\bibfnamefont {B.-S.}\ \bibnamefont {Zou}},\
  }\href {\doibase 10.1103/PhysRevD.104.114001} {\bibfield  {journal} {\bibinfo
   {journal} {Phys. Rev. D}\ }\textbf {\bibinfo {volume} {104}},\ \bibinfo
  {pages} {114001} (\bibinfo {year} {2021})},\ \Eprint
  {http://arxiv.org/abs/2107.14470} {arXiv:2107.14470 [hep-ph]} \BibitemShut
  {NoStop}%
\bibitem [{\citenamefont {Nagahiro}\ \emph {et~al.}(2008)\citenamefont
  {Nagahiro}, \citenamefont {Roca}, \citenamefont {Oset},\ and\ \citenamefont
  {Zou}}]{Nagahiro:2008bn}%
  \BibitemOpen
  \bibfield  {author} {\bibinfo {author} {\bibfnamefont {H.}~\bibnamefont
  {Nagahiro}}, \bibinfo {author} {\bibfnamefont {L.}~\bibnamefont {Roca}},
  \bibinfo {author} {\bibfnamefont {E.}~\bibnamefont {Oset}}, \ and\ \bibinfo
  {author} {\bibfnamefont {B.~S.}\ \bibnamefont {Zou}},\ }\href {\doibase
  10.1103/PhysRevD.78.014012} {\bibfield  {journal} {\bibinfo  {journal} {Phys.
  Rev. D}\ }\textbf {\bibinfo {volume} {78}},\ \bibinfo {pages} {014012}
  (\bibinfo {year} {2008})},\ \Eprint {http://arxiv.org/abs/0803.4460}
  {arXiv:0803.4460 [hep-ph]} \BibitemShut {NoStop}%
\bibitem [{\citenamefont {Branz}\ \emph {et~al.}(2010)\citenamefont {Branz},
  \citenamefont {Geng},\ and\ \citenamefont {Oset}}]{Branz:2009cv}%
  \BibitemOpen
  \bibfield  {author} {\bibinfo {author} {\bibfnamefont {T.}~\bibnamefont
  {Branz}}, \bibinfo {author} {\bibfnamefont {L.~S.}\ \bibnamefont {Geng}}, \
  and\ \bibinfo {author} {\bibfnamefont {E.}~\bibnamefont {Oset}},\ }\href
  {\doibase 10.1103/PhysRevD.81.054037} {\bibfield  {journal} {\bibinfo
  {journal} {Phys. Rev. D}\ }\textbf {\bibinfo {volume} {81}},\ \bibinfo
  {pages} {054037} (\bibinfo {year} {2010})},\ \Eprint
  {http://arxiv.org/abs/0911.0206} {arXiv:0911.0206 [hep-ph]} \BibitemShut
  {NoStop}%
\bibitem [{\citenamefont {Geng}\ \emph {et~al.}()\citenamefont {Geng},
  \citenamefont {Guo}, \citenamefont {Hanhart}, \citenamefont {Molina},
  \citenamefont {Oset},\ and\ \citenamefont {Zou}}]{Geng:2010kma}%
  \BibitemOpen
  \bibfield  {author} {\bibinfo {author} {\bibfnamefont {L.~S.}\ \bibnamefont
  {Geng}}, \bibinfo {author} {\bibfnamefont {F.~K.}\ \bibnamefont {Guo}},
  \bibinfo {author} {\bibfnamefont {C.}~\bibnamefont {Hanhart}}, \bibinfo
  {author} {\bibfnamefont {R.}~\bibnamefont {Molina}}, \bibinfo {author}
  {\bibfnamefont {E.}~\bibnamefont {Oset}}, \ and\ \bibinfo {author}
  {\bibfnamefont {B.~S.}\ \bibnamefont {Zou}},\ }\href {\doibase
  10.1140/epja/i2010-10971-5} {\bibfield  {journal} {\bibinfo  {journal} {Eur.
  Phys. J. A}\ }\textbf {\bibinfo {volume} {44}},\ \bibinfo {pages} {305}},\
  \Eprint {http://arxiv.org/abs/0910.5192} {arXiv:0910.5192 [hep-ph]}
  \BibitemShut {NoStop}%
\bibitem [{\citenamefont {Martinez~Torres}\ \emph {et~al.}(2013)\citenamefont
  {Martinez~Torres}, \citenamefont {Khemchandani}, \citenamefont {Navarra},
  \citenamefont {Nielsen},\ and\ \citenamefont {Oset}}]{MartinezTorres:2012du}%
  \BibitemOpen
  \bibfield  {author} {\bibinfo {author} {\bibfnamefont {A.}~\bibnamefont
  {Martinez~Torres}}, \bibinfo {author} {\bibfnamefont {K.~P.}\ \bibnamefont
  {Khemchandani}}, \bibinfo {author} {\bibfnamefont {F.~S.}\ \bibnamefont
  {Navarra}}, \bibinfo {author} {\bibfnamefont {M.}~\bibnamefont {Nielsen}}, \
  and\ \bibinfo {author} {\bibfnamefont {E.}~\bibnamefont {Oset}},\ }\href
  {\doibase 10.1016/j.physletb.2013.01.036} {\bibfield  {journal} {\bibinfo
  {journal} {Phys. Lett. B}\ }\textbf {\bibinfo {volume} {719}},\ \bibinfo
  {pages} {388} (\bibinfo {year} {2013})},\ \Eprint
  {http://arxiv.org/abs/1210.6392} {arXiv:1210.6392 [hep-ph]} \BibitemShut
  {NoStop}%
\bibitem [{\citenamefont {Xie}\ and\ \citenamefont {Oset}(2014)}]{Xie:2014gla}%
  \BibitemOpen
  \bibfield  {author} {\bibinfo {author} {\bibfnamefont {J.-J.}\ \bibnamefont
  {Xie}}\ and\ \bibinfo {author} {\bibfnamefont {E.}~\bibnamefont {Oset}},\
  }\href {\doibase 10.1103/PhysRevD.90.094006} {\bibfield  {journal} {\bibinfo
  {journal} {Phys. Rev. D}\ }\textbf {\bibinfo {volume} {90}},\ \bibinfo
  {pages} {094006} (\bibinfo {year} {2014})},\ \Eprint
  {http://arxiv.org/abs/1409.1341} {arXiv:1409.1341 [hep-ph]} \BibitemShut
  {NoStop}%
\bibitem [{\citenamefont {Dai}\ \emph {et~al.}(2015)\citenamefont {Dai},
  \citenamefont {Xie},\ and\ \citenamefont {Oset}}]{Dai:2015cwa}%
  \BibitemOpen
  \bibfield  {author} {\bibinfo {author} {\bibfnamefont {L.-R.}\ \bibnamefont
  {Dai}}, \bibinfo {author} {\bibfnamefont {J.-J.}\ \bibnamefont {Xie}}, \ and\
  \bibinfo {author} {\bibfnamefont {E.}~\bibnamefont {Oset}},\ }\href {\doibase
  10.1103/PhysRevD.91.094013} {\bibfield  {journal} {\bibinfo  {journal} {Phys.
  Rev. D}\ }\textbf {\bibinfo {volume} {91}},\ \bibinfo {pages} {094013}
  (\bibinfo {year} {2015})},\ \Eprint {http://arxiv.org/abs/1503.02463}
  {arXiv:1503.02463 [hep-ph]} \BibitemShut {NoStop}%
\bibitem [{\citenamefont {Molina}\ \emph {et~al.}(2020)\citenamefont {Molina},
  \citenamefont {Dai}, \citenamefont {Geng},\ and\ \citenamefont
  {Oset}}]{Molina:2019wjj}%
  \BibitemOpen
  \bibfield  {author} {\bibinfo {author} {\bibfnamefont {R.}~\bibnamefont
  {Molina}}, \bibinfo {author} {\bibfnamefont {L.~R.}\ \bibnamefont {Dai}},
  \bibinfo {author} {\bibfnamefont {L.~S.}\ \bibnamefont {Geng}}, \ and\
  \bibinfo {author} {\bibfnamefont {E.}~\bibnamefont {Oset}},\ }\href {\doibase
  10.1140/epja/s10050-020-00176-y} {\bibfield  {journal} {\bibinfo  {journal}
  {Eur. Phys. J. A}\ }\textbf {\bibinfo {volume} {56}},\ \bibinfo {pages} {173}
  (\bibinfo {year} {2020})},\ \Eprint {http://arxiv.org/abs/1909.10764}
  {arXiv:1909.10764 [hep-ph]} \BibitemShut {NoStop}%
\bibitem [{\citenamefont {Roca}\ and\ \citenamefont
  {Oset}(2010)}]{Roca:2010tf}%
  \BibitemOpen
  \bibfield  {author} {\bibinfo {author} {\bibfnamefont {L.}~\bibnamefont
  {Roca}}\ and\ \bibinfo {author} {\bibfnamefont {E.}~\bibnamefont {Oset}},\
  }\href {\doibase 10.1103/PhysRevD.82.054013} {\bibfield  {journal} {\bibinfo
  {journal} {Phys. Rev. D}\ }\textbf {\bibinfo {volume} {82}},\ \bibinfo
  {pages} {054013} (\bibinfo {year} {2010})},\ \Eprint
  {http://arxiv.org/abs/1005.0283} {arXiv:1005.0283 [hep-ph]} \BibitemShut
  {NoStop}%
\bibitem [{\citenamefont {Gamermann}\ \emph {et~al.}(2010)\citenamefont
  {Gamermann}, \citenamefont {Nieves}, \citenamefont {Oset},\ and\
  \citenamefont {Ruiz~Arriola}}]{Gamermann:2009uq}%
  \BibitemOpen
  \bibfield  {author} {\bibinfo {author} {\bibfnamefont {D.}~\bibnamefont
  {Gamermann}}, \bibinfo {author} {\bibfnamefont {J.}~\bibnamefont {Nieves}},
  \bibinfo {author} {\bibfnamefont {E.}~\bibnamefont {Oset}}, \ and\ \bibinfo
  {author} {\bibfnamefont {E.}~\bibnamefont {Ruiz~Arriola}},\ }\href {\doibase
  10.1103/PhysRevD.81.014029} {\bibfield  {journal} {\bibinfo  {journal} {Phys.
  Rev. D}\ }\textbf {\bibinfo {volume} {81}},\ \bibinfo {pages} {014029}
  (\bibinfo {year} {2010})},\ \Eprint {http://arxiv.org/abs/0911.4407}
  {arXiv:0911.4407 [hep-ph]} \BibitemShut {NoStop}%
\bibitem [{\citenamefont {Yamagata-Sekihara}\ \emph {et~al.}(2011)\citenamefont
  {Yamagata-Sekihara}, \citenamefont {Nieves},\ and\ \citenamefont
  {Oset}}]{Yamagata-Sekihara:2010kpd}%
  \BibitemOpen
  \bibfield  {author} {\bibinfo {author} {\bibfnamefont {J.}~\bibnamefont
  {Yamagata-Sekihara}}, \bibinfo {author} {\bibfnamefont {J.}~\bibnamefont
  {Nieves}}, \ and\ \bibinfo {author} {\bibfnamefont {E.}~\bibnamefont
  {Oset}},\ }\href {\doibase 10.1103/PhysRevD.83.014003} {\bibfield  {journal}
  {\bibinfo  {journal} {Phys. Rev. D}\ }\textbf {\bibinfo {volume} {83}},\
  \bibinfo {pages} {014003} (\bibinfo {year} {2011})},\ \Eprint
  {http://arxiv.org/abs/1007.3923} {arXiv:1007.3923 [hep-ph]} \BibitemShut
  {NoStop}%
\bibitem [{\citenamefont {Zyla}\ \emph {et~al.}(2020)\citenamefont {Zyla} \emph
  {et~al.}}]{ParticleDataGroup:2020ssz}%
  \BibitemOpen
  \bibfield  {author} {\bibinfo {author} {\bibfnamefont {P.~A.}\ \bibnamefont
  {Zyla}} \emph {et~al.} (\bibinfo {collaboration} {Particle Data Group}),\
  }\href {\doibase 10.1093/ptep/ptaa104} {\bibfield  {journal} {\bibinfo
  {journal} {PTEP}\ }\textbf {\bibinfo {volume} {2020}},\ \bibinfo {pages}
  {083C01} (\bibinfo {year} {2020})}\BibitemShut {NoStop}%
\bibitem [{\citenamefont {Zhu}\ \emph {et~al.}(2022)\citenamefont {Zhu},
  \citenamefont {Li}, \citenamefont {Wang}, \citenamefont {Geng},\ and\
  \citenamefont {Xie}}]{Zhu:2022wzk}%
  \BibitemOpen
  \bibfield  {author} {\bibinfo {author} {\bibfnamefont {X.}~\bibnamefont
  {Zhu}}, \bibinfo {author} {\bibfnamefont {D.-M.}\ \bibnamefont {Li}},
  \bibinfo {author} {\bibfnamefont {E.}~\bibnamefont {Wang}}, \bibinfo {author}
  {\bibfnamefont {L.-S.}\ \bibnamefont {Geng}}, \ and\ \bibinfo {author}
  {\bibfnamefont {J.-J.}\ \bibnamefont {Xie}},\ }\href {\doibase
  10.1103/PhysRevD.105.116010} {\bibfield  {journal} {\bibinfo  {journal}
  {Phys. Rev. D}\ }\textbf {\bibinfo {volume} {105}},\ \bibinfo {pages}
  {116010} (\bibinfo {year} {2022})},\ \Eprint
  {http://arxiv.org/abs/2204.09384} {arXiv:2204.09384 [hep-ph]} \BibitemShut
  {NoStop}%
\bibitem [{\citenamefont {Geng}\ \emph {et~al.}(2007)\citenamefont {Geng},
  \citenamefont {Oset}, \citenamefont {Roca},\ and\ \citenamefont
  {Oller}}]{Geng:2006yb}%
  \BibitemOpen
  \bibfield  {author} {\bibinfo {author} {\bibfnamefont {L.~S.}\ \bibnamefont
  {Geng}}, \bibinfo {author} {\bibfnamefont {E.}~\bibnamefont {Oset}}, \bibinfo
  {author} {\bibfnamefont {L.}~\bibnamefont {Roca}}, \ and\ \bibinfo {author}
  {\bibfnamefont {J.~A.}\ \bibnamefont {Oller}},\ }\href {\doibase
  10.1103/PhysRevD.75.014017} {\bibfield  {journal} {\bibinfo  {journal} {Phys.
  Rev. D}\ }\textbf {\bibinfo {volume} {75}},\ \bibinfo {pages} {014017}
  (\bibinfo {year} {2007})},\ \Eprint {http://arxiv.org/abs/hep-ph/0610217}
  {arXiv:hep-ph/0610217} \BibitemShut {NoStop}%
\bibitem [{\citenamefont {Guo}\ \emph {et~al.}(2006)\citenamefont {Guo},
  \citenamefont {Ping}, \citenamefont {Shen}, \citenamefont {Chiang},\ and\
  \citenamefont {Zou}}]{Guo:2005wp}%
  \BibitemOpen
  \bibfield  {author} {\bibinfo {author} {\bibfnamefont {F.-K.}\ \bibnamefont
  {Guo}}, \bibinfo {author} {\bibfnamefont {R.-G.}\ \bibnamefont {Ping}},
  \bibinfo {author} {\bibfnamefont {P.-N.}\ \bibnamefont {Shen}}, \bibinfo
  {author} {\bibfnamefont {H.-C.}\ \bibnamefont {Chiang}}, \ and\ \bibinfo
  {author} {\bibfnamefont {B.-S.}\ \bibnamefont {Zou}},\ }\href {\doibase
  10.1016/j.nuclphysa.2006.04.008} {\bibfield  {journal} {\bibinfo  {journal}
  {Nucl. Phys. A}\ }\textbf {\bibinfo {volume} {773}},\ \bibinfo {pages} {78}
  (\bibinfo {year} {2006})},\ \Eprint {http://arxiv.org/abs/hep-ph/0509050}
  {arXiv:hep-ph/0509050} \BibitemShut {NoStop}%
\bibitem [{\citenamefont {Sun}\ \emph {et~al.}(2022)\citenamefont {Sun},
  \citenamefont {Fan},\ and\ \citenamefont {Cao}}]{Sun:2022cxf}%
  \BibitemOpen
  \bibfield  {author} {\bibinfo {author} {\bibfnamefont {B.-X.}\ \bibnamefont
  {Sun}}, \bibinfo {author} {\bibfnamefont {Y.-Y.}\ \bibnamefont {Fan}}, \ and\
  \bibinfo {author} {\bibfnamefont {Q.-Q.}\ \bibnamefont {Cao}},\ }\href@noop
  {} {\  (\bibinfo {year} {2022})},\ \Eprint {http://arxiv.org/abs/2206.02961}
  {arXiv:2206.02961 [hep-ph]} \BibitemShut {NoStop}%
\bibitem [{\citenamefont {Bisello}\ \emph {et~al.}(1989)\citenamefont {Bisello}
  \emph {et~al.}}]{DM2:1988esg}%
  \BibitemOpen
  \bibfield  {author} {\bibinfo {author} {\bibfnamefont {D.}~\bibnamefont
  {Bisello}} \emph {et~al.} (\bibinfo {collaboration} {DM2}),\ }\href {\doibase
  10.1103/PhysRevD.39.701} {\bibfield  {journal} {\bibinfo  {journal} {Phys.
  Rev. D}\ }\textbf {\bibinfo {volume} {39}},\ \bibinfo {pages} {701} (\bibinfo
  {year} {1989})}\BibitemShut {NoStop}%
\bibitem [{\citenamefont {Ablikim}\ \emph {et~al.}(2016)\citenamefont {Ablikim}
  \emph {et~al.}}]{BESIII:2016qzq}%
  \BibitemOpen
  \bibfield  {author} {\bibinfo {author} {\bibfnamefont {M.}~\bibnamefont
  {Ablikim}} \emph {et~al.} (\bibinfo {collaboration} {BESIII}),\ }\href
  {\doibase 10.1103/PhysRevD.93.112011} {\bibfield  {journal} {\bibinfo
  {journal} {Phys. Rev. D}\ }\textbf {\bibinfo {volume} {93}},\ \bibinfo
  {pages} {112011} (\bibinfo {year} {2016})},\ \Eprint
  {http://arxiv.org/abs/1602.01523} {arXiv:1602.01523 [hep-ex]} \BibitemShut
  {NoStop}%
\bibitem [{\citenamefont {Wang}\ \emph {et~al.}(2017)\citenamefont {Wang},
  \citenamefont {Luo}, \citenamefont {Sun},\ and\ \citenamefont
  {Liu}}]{Wang:2017iai}%
  \BibitemOpen
  \bibfield  {author} {\bibinfo {author} {\bibfnamefont {L.-M.}\ \bibnamefont
  {Wang}}, \bibinfo {author} {\bibfnamefont {S.-Q.}\ \bibnamefont {Luo}},
  \bibinfo {author} {\bibfnamefont {Z.-F.}\ \bibnamefont {Sun}}, \ and\
  \bibinfo {author} {\bibfnamefont {X.}~\bibnamefont {Liu}},\ }\href {\doibase
  10.1103/PhysRevD.96.034013} {\bibfield  {journal} {\bibinfo  {journal} {Phys.
  Rev. D}\ }\textbf {\bibinfo {volume} {96}},\ \bibinfo {pages} {034013}
  (\bibinfo {year} {2017})},\ \Eprint {http://arxiv.org/abs/1705.00549}
  {arXiv:1705.00549 [hep-ph]} \BibitemShut {NoStop}%
\bibitem [{\citenamefont {Anisovich}\ \emph {et~al.}(2001)\citenamefont
  {Anisovich}, \citenamefont {Baker}, \citenamefont {Batty}, \citenamefont
  {Bugg}, \citenamefont {Nikonov}, \citenamefont {Sarantsev}, \citenamefont
  {Sarantsev},\ and\ \citenamefont {Zou}}]{Anisovich:2001pn}%
  \BibitemOpen
  \bibfield  {author} {\bibinfo {author} {\bibfnamefont {A.~V.}\ \bibnamefont
  {Anisovich}}, \bibinfo {author} {\bibfnamefont {C.~A.}\ \bibnamefont
  {Baker}}, \bibinfo {author} {\bibfnamefont {C.~J.}\ \bibnamefont {Batty}},
  \bibinfo {author} {\bibfnamefont {D.~V.}\ \bibnamefont {Bugg}}, \bibinfo
  {author} {\bibfnamefont {V.~A.}\ \bibnamefont {Nikonov}}, \bibinfo {author}
  {\bibfnamefont {A.~V.}\ \bibnamefont {Sarantsev}}, \bibinfo {author}
  {\bibfnamefont {V.~V.}\ \bibnamefont {Sarantsev}}, \ and\ \bibinfo {author}
  {\bibfnamefont {B.~S.}\ \bibnamefont {Zou}},\ }\href {\doibase
  10.1016/S0370-2693(01)01017-6} {\bibfield  {journal} {\bibinfo  {journal}
  {Phys. Lett. B}\ }\textbf {\bibinfo {volume} {517}},\ \bibinfo {pages} {261}
  (\bibinfo {year} {2001})},\ \Eprint {http://arxiv.org/abs/1110.0278}
  {arXiv:1110.0278 [hep-ex]} \BibitemShut {NoStop}%
\bibitem [{\citenamefont {Ablikim}\ \emph {et~al.}(2004)\citenamefont {Ablikim}
  \emph {et~al.}}]{BES:2004fgd}%
  \BibitemOpen
  \bibfield  {author} {\bibinfo {author} {\bibfnamefont {M.}~\bibnamefont
  {Ablikim}} \emph {et~al.} (\bibinfo {collaboration} {BES}),\ }\href {\doibase
  10.1103/PhysRevLett.93.112002} {\bibfield  {journal} {\bibinfo  {journal}
  {Phys. Rev. Lett.}\ }\textbf {\bibinfo {volume} {93}},\ \bibinfo {pages}
  {112002} (\bibinfo {year} {2004})},\ \Eprint
  {http://arxiv.org/abs/hep-ex/0405050} {arXiv:hep-ex/0405050} \BibitemShut
  {NoStop}%
\bibitem [{\citenamefont {Chen}\ \emph {et~al.}(2011)\citenamefont {Chen} \emph
  {et~al.}}]{Belle:2011cxw}%
  \BibitemOpen
  \bibfield  {author} {\bibinfo {author} {\bibfnamefont {P.}~\bibnamefont
  {Chen}} \emph {et~al.} (\bibinfo {collaboration} {Belle}),\ }\href {\doibase
  10.1103/PhysRevD.84.071501} {\bibfield  {journal} {\bibinfo  {journal} {Phys.
  Rev. D}\ }\textbf {\bibinfo {volume} {84}},\ \bibinfo {pages} {071501}
  (\bibinfo {year} {2011})},\ \Eprint {http://arxiv.org/abs/1108.4271}
  {arXiv:1108.4271 [hep-ex]} \BibitemShut {NoStop}%
\end{thebibliography}%

\end{document}